\documentclass[a4paper,fleqn]{cas-dc}
\usepackage[T1]{fontenc}
\usepackage[utf8]{inputenc}
\usepackage{graphicx}
\usepackage{subcaption}
\usepackage{comment}
\usepackage{tabularx}
\usepackage{booktabs}
\usepackage{rotating}
\usepackage{amsmath}
\usepackage{xurl}
\usepackage{amssymb}
\usepackage{algorithm, algpseudocode}
\usepackage{tikz}
\usepackage[section]{placeins}
\usepackage{natbib}
\usepackage{xcolor}
\usetikzlibrary{shapes.geometric, arrows, calc, positioning}

\tikzstyle{process} = [rectangle, rounded corners, minimum width=3.2cm, minimum height=1cm, text centered, draw=black]
\tikzstyle{decision} = [diamond, minimum width=3cm, minimum height=1cm, text centered, draw=black]

\begin{document}
\let\WriteBookmarks\relax
\def\floatpagepagefraction{1}
\def\textpagefraction{.001}

\shorttitle{Progressive Lower Bound Evaluation for Urban Road Renovation Scheduling}
\shortauthors{Bosch et~al.}

\title [mode = title]{Progressive Lower Bound Evaluation for Urban Road Renovation Scheduling}

\author[inst1]{Robbert Bosch}[orcid=0009-0006-8807-1031]
\author[inst1]{Patricia Rogetzer}[orcid=0000-0001-9582-6320]
\author[inst1]{Wouter van Heeswijk}[orcid=0000-0002-5413-9660]
\author[inst1]{Martijn Mes}[orcid=0000-0001-9676-5259]

\affiliation[inst1]{organization={University of Twente}, 
            addressline={P.O. Box 217}, 
            city={Enschede},
            postcode={7500 AE}, 
            country={The Netherlands}}

\begin{abstract}
Urban infrastructure degrades over time, necessitating periodic renovation to maintain functionality and safety. When renovation is delayed beyond the infrastructure's remaining lifespan, costly emergency interventions become necessary to prevent failure. Decision makers must therefore balance expected emergency intervention costs against traffic congestion impacts. We formalize this trade-off as a road network maintenance scheduling problem with uncertain deadlines, which presents optimization challenges including computationally expensive evaluation and an exponentially growing solution space. To address these challenges, this paper contributes a hybrid optimization approach combining machine learning with genetic algorithms for large-scale infrastructure renovation scheduling under uncertainty. We formulate the problem as a bi-level multi-objective optimization problem that explicitly accounts for uncertain infrastructure lifespans through probabilistic failure models. We develop a progressive lower bound evaluation method: candidate schedules receive quick, optimistic estimates of their traffic impact from machine learning based surrogate models, and these estimates are refined by exact traffic simulation only for promising candidates, so that full evaluation is reserved almost exclusively for schedules that can advance the Pareto front. Embedded in a multi-objective genetic algorithm, this evaluation strategy improves solution quality by enabling many more iterations within a fixed computational budget. We demonstrate the method's effectiveness and scalability on substantially larger problem instances than previously addressed in the literature: 76 and 80 projects on the Sioux Falls and Anaheim networks, respectively. Under an identical 24-hour computational budget across 30 random seeds, the method completes up to 87 times as many optimization iterations as standard NSGA-II and achieves statistically significant improvements across multiple performance metrics over both standard NSGA-II and a surrogate-assisted baseline. For road agencies, these gains make city-scale renovation programming tractable within a single day of computation: planners obtain a Pareto front of implementable schedules that quantifies the trade-off between the risk of late renovations and network congestion, from which a preferred schedule can be selected and defended, or adjustments explored with their consequences quantified.
\end{abstract}


\maketitle

\section{Introduction}
\label{sec:introduction}
The planning and scheduling of urban infrastructure renovation projects pose a complex decision-making problem with increasing practical relevance in modern day cities. Infrastructure such as roads, bridges, and quay walls degrade over time, requiring periodic renovation to maintain functionality and safety. However, remaining lifespan is inherently uncertain due to nonlinear degradation processes, limited inspection capabilities \citep{biondiniLifeCyclePerformanceDeteriorating2016} and sometimes historical character of infrastructure. Recent failures, such as the unexpected collapse of quay walls in Amsterdam's city center \citep{nosKademuurCentrumAmsterdam2020}, highlight the need for proactive scheduling approaches. Throughout this paper, ``failure'' denotes infrastructure reaching a critical state that requires emergency intervention (such as temporary support structures) rather than literal collapse; such measures prevent catastrophic outcomes but impose substantial costs that effective scheduling aims to avoid.

Urban infrastructure renovations impose substantial costs: they are expensive, time-consuming, and most critically, disrupt local traffic networks. Decision makers face a fundamental trade-off. Scheduling renovations early reduces the risk of infrastructure reaching critical condition—which necessitates costly and disruptive emergency interventions such as temporary support structures—but increases traffic disruption duration. Conversely, delaying renovations minimizes immediate congestion but increases the likelihood of such emergency measures. Critically, the increase in traffic congestion depends on which projects operate simultaneously—closing two major arterial roads concurrently can cause significantly more or less congestion than staggering these closures due to complex traffic flow interactions \citep{leeOptimizingScheduleImproving2009}. This interdependency necessitates coordinated scheduling rather than project-by-project planning.

Road renovation affects a broad set of stakeholders beyond drivers. Pedestrians and other vulnerable users are concerned with safety, local businesses and community facilities with continued access \citep{miralinaghiOptimizingSelectionScheduling2020, miralinaghiNetworklevelSchedulingRoad2020}, residents with construction-related noise, dust, and air pollution \citep{alvanchiConstructionScheduleInfluential2020, kimEnvironmentalImpactsTraffic2024}, and the responsible agency with life-cycle maintenance and emission costs \citep{seilabiReinforcementLearningbasedApproach2025, chowdhuryMultiobjectiveOptimizationMaintenance2025}. Indeed, road-work scheduling decisions have been shown to influence a broad spectrum of outcomes spanning safety, mobility, economic, and environmental impacts \citep{gongOptimizingSchedulingLongterm2016}. Jointly modeling all of these concerns is beyond the scope of a single tractable formulation; we therefore focus on the two objectives most directly coupled through the road network---expected failure cost and aggregate travel delay---while recognizing this broader objective space, to which we return in Section~\ref{sec:conclusion}.

The problem of optimally scheduling urban infrastructure renovations to balance failure risk with traffic congestion, which we term the Road Network Maintenance Scheduling Problem with Uncertain Deadlines (RNMSP-UD), presents three key optimization challenges. First, it involves multiple conflicting objectives, requiring multi-objective optimization to generate trade-off solutions between emergency intervention costs and traffic disruption. Second, it exhibits a bi-level structure where upper level scheduling decisions determine network capacity while lower level driver routing decisions determine actual congestion. The actual congestion emerges from collective routing decisions that can only be captured through traffic simulations for each discrete time period in the planning horizon, which makes evaluation of schedules computationally expensive. Third, the solution space for scheduling problems grows exponentially with the problem size: For a set of projects $\mathcal{P}$ and a set of feasible starting times $\mathcal{T}$, there are  $|\mathcal{T}|^{|\mathcal{P}|}$ potential schedules, disregarding scheduling constraints such as budget or construction capacity.

Despite practical importance, existing literature exhibits two key limitations. First, most research assumes fixed renovation deadlines and does not model uncertain infrastructure lifespans or explore trade-offs between failure risk and traffic impacts under uncertainty. Second, scalability remains limited---existing methods are typically evaluated on small case studies with fewer than 35 projects (see Table~\ref{table:lit_overview}), raising concerns about applicability to real-world urban networks requiring coordination of dozens or hundreds of projects.

To address the scalability issues, this paper contributes a hybrid optimization approach combining machine learning with genetic algorithms to enable large-scale infrastructure renovation scheduling under uncertainty. Our approach, termed Progressive Lower Bound Evaluation (PLBE), targets the traffic simulations, which are by a large margin the most computationally intensive component of evaluating a schedule: a schedule's total travel delay is a sum of per-period traffic-scenario delays, each distinct scenario requiring one simulation. PLBE evaluates a candidate in a stepwise manner: machine learning based surrogate models supply deliberately optimistic estimates for the not-yet-simulated scenarios, so that the combination of exact and estimated delays constitutes a conservative lower-bound estimate of the schedule's total delay. If this bound is already dominated by schedules found so far, the candidate is discarded; otherwise, one scenario is simulated exactly and the bound is tightened---repeating until the candidate is either discarded or fully evaluated. Without such a bound, an optimizer must either simulate every candidate fully---prohibitive at realistic problem sizes---or rely on a point prediction of the complete objective, which cannot tell confidently poor candidates from mispredicted good ones and therefore risks discarding genuinely promising schedules (a failure mode we quantify empirically in Section~\ref{subsec:exp1_new}). PLBE thus drastically reduces computational requirements while preserving solution quality; Section~\ref{subsec: related optimization} contrasts it with conventional surrogate-assisted algorithms and Section~\ref{sec:solmethod} details the mechanism.

This paper makes three key contributions. First, we formulate the RNMSP-UD as a bi-level multi-objective optimization problem that explicitly models uncertain infrastructure lifespans through probabilistic failure models, introducing failure risk as an objective alongside traffic disruption. While multi-objective formulations exist in road work scheduling literature, this specific objective is novel. The multi-objective nature—which inherently requires finding a set of optimal solutions rather than a single optimum—combined with the bi-level structure and expensive traffic simulations, creates a critical scalability challenge. Second, to address this computational bottleneck, we develop PLBE, a method that integrates surrogate models that predict traffic simulation outcomes into Non-Dominated Sorting Genetic Algorithm II (NSGA-II), a Multi-Objective Genetic Algorithm, to prune suboptimal solutions, reducing the number of required traffic simulations. Third, we demonstrate PLBE's effectiveness on the Sioux Falls network with 76 projects—substantially larger than the fewer than 35 projects typical in similar studies—achieving up to an 87$\times$ increase in optimization iterations completed within the same time budget compared to standard NSGA-II---our control baseline, run with identical operators, population, and a shared 24-hour time budget---while improving solution quality, with statistical significance assessed across 30 random seeds using non-parametric tests; on the larger Anaheim network (80 projects), PLBE's Pareto front fully dominates the baseline's.

The remainder of this paper is structured as follows. Section~\ref{sec:literature} reviews existing research and identifies gaps. Sections~\ref{sec:problem} and \ref{sec:solmethod} present our problem formulation and solution algorithm, respectively. Section~\ref{sec:experiments} provides experimental validation on the Sioux Falls and Anaheim networks, and Section~\ref{sec:conclusion} concludes with implications for practice and future research.

\section{Literature}
\label{sec:literature}

In this section, we analyze existing approaches to infrastructure renovation project scheduling, highlighting their limitations and positioning our contribution. Section \ref{subsec: related problem classes} examines related problem classes and their key characteristics. Section \ref{subsec: related optimization} evaluates common optimization methods and their computational challenges. Section \ref{subsec: literature gap} identifies critical gaps that motivate our approach.

\subsection{Related Problem Classes}
\label{subsec: related problem classes}

Road work scheduling problems (RWSPs) involve determining when to initiate projects, with the core decision being the starting time for each project, sometimes augmented with secondary decisions. We classify these problems into three categories based on scope and objectives. Table \ref{tab:problem-overview} provides an overview of related problem classes, which we expand upon in this section.

\begin{table*}
\centering
\caption{Overview of related problem classes}
\label{tab:problem-overview}
\small
\begin{tabularx}{\textwidth}{p{1.1cm} X p{1.2cm} p{1.2cm} p{1.7cm}}     
\toprule
Problem & Goal & Physical scale & Temporal scale & Permanent Changes \\
\midrule
NDP & Optimal design of traffic network through addition of new roads or road expansions. & City & Years & Yes \\
WZSP & Minimize hindrance experienced on a large traffic artery due to road works. & Road & Weeks & No \\
RNMSP & Minimize hindrance of local traffic caused by a set of possibly simultaneous road works. & City & Years & No \\
\bottomrule
\end{tabularx}
\end{table*}

Network Design Problems (NDPs) focus on expanding transportation capacity through new construction or major upgrades. The primary decision involves both selecting which projects to build and scheduling their implementation to maximize long-term network improvements. Objectives typically center on maximizing travel time savings \citep{poorzahedyApplicationAntSystem2005} or broader economic benefits like consumer surplus. Some studies consider only project selection \citep{farahaniReviewUrbanTransportation2013, yamanyNetworklevelPavementMaintenance2024}, while others integrate selection with scheduling to minimize construction-phase disruptions \citep{ukkusuriMultiperiodTransportationNetwork2009, taoIslandModelsStochastic2007, hosseininasabIntegrationSelectingScheduling2015}. Multi-objective approaches often balance network performance against secondary impacts such as land use changes \citep{szetoTimeDependentDiscreteNetwork2010}, public health \citep{jiangTimedependentTransportationNetwork2015}, or equity considerations \citep{hosseininasabMultiobjectiveIntegratedModel2018}.

NDPs differ fundamentally from maintenance scheduling in two ways. First, because NDPs consider selection and/or scheduling of construction projects rather than renovation projects, the result of such operations permanently changes the traffic network. This changes the focus from minimizing temporary disruptions to maximizing long-term gains. Second, construction projects rarely face hard deadlines since delays, while costly, do not risk catastrophic infrastructure failure. This contrasts sharply with maintenance projects where postponement increases collapse risk.

Work Zone Scheduling Problems (WZSPs) address the maintenance of individual highway segments, optimizing both spatial aspects such as lane closure configurations and work zone length as well as temporal scheduling. The scheduling decisions operate at a tactical level, determining optimal work hours and phasing to minimize congestion on the specific roadway. These problems typically employ microscopic traffic simulation to capture detailed traffic flow dynamics \citep{changTabuSearchBased} and may consider traffic diversion effects \citep{chienSchedulingHighwayWork2014}. However, by focusing on individual roads, WZSPs cannot capture the network-wide interactions that occur when multiple projects operate simultaneously across different parts of an urban network.

Road Network Maintenance Scheduling Problems (RNMSPs) operate at the network level, coordinating multiple maintenance projects across an urban or regional road system. This is the problem class addressed in our paper. Like NDPs, RNMSPs require network-wide traffic impact assessment since interactions between simultaneous road closures significantly influence total system congestion; like WZSPs, they focus on maintenance activities rather than capacity expansion.
The RNMSP literature has evolved to incorporate increasingly realistic modeling assumptions. Early work established the basic framework of minimizing network-wide travel delays through coordinated scheduling \citep{cheuGeneticAlgorithmSimulationMethodology2004, gongOptimizingSchedulingLongterm2016}. Subsequent research has extended this foundation in several directions: some studies model how drivers adapt their routes in response to closures rather than assuming fixed routing patterns \citep{leeOptimizingScheduleImproving2009}, while others allow travel demand itself to vary with network conditions \citep{liBilevelOptimizationLongterm2021}. The types of closures considered have also expanded from simple binary road removals to partial capacity reductions and lane-specific restrictions \citep{seilabiReinforcementLearningbasedApproach2025}. Beyond routine maintenance, some work addresses emergency scenarios such as post-disaster repair sequencing, where urgency and network connectivity introduce additional constraints \citep{moghtadernejadPrioritizingRoadNetwork2022}. Some studies incorporate objectives beyond traffic performance, such as the financial implications of contractor bundling arrangements \citep{miralinaghiContractBundlingConsiderations2022}. A notable gap across this literature concerns the treatment of project deadlines. Existing approaches typically assume fixed deadlines or omit them entirely, yet infrastructure lifespans are inherently uncertain. To our knowledge, no existing work explicitly models this uncertainty or explores the resulting trade-off between scheduling early to reduce failure risk and scheduling late to minimize traffic disruption.

\subsection{Common Optimization Methods for RWSPs}
\label{subsec: related optimization}

Road work scheduling problems share three characteristics that make optimization challenging: exponential solution spaces, expensive evaluations requiring traffic simulations, and multiple conflicting objectives.

As mentioned in Section~\ref{sec:introduction}, the solution space grows exponentially with respect to the number of projects and feasible starting times per project. More critically, these problems exhibit a bi-level structure: upper level scheduling decisions affect network capacity throughout the planning horizon, while lower level driver routing decisions determine actual congestion levels. This bi-level nature means that each schedule evaluation requires solving a Traffic Assignment Problem (TAP) for multiple time periods, with individual traffic simulations taking seconds to hours depending on network size and level of detail \citep{chenRoadsideLiDARPlacement2024}. The expensive black-box nature of schedule evaluations creates a scalability bottleneck. Unlike problems where objective functions can be evaluated analytically, traffic congestion emerges from complex driver behavior that can only be captured through simulation. This makes gradient-based methods inapplicable and limits the range of available optimization approaches.

Heuristic methods such as cost-benefit analysis \citep{kumarSimplifiedFrameworkSequencing2018} and greedy algorithms are computationally efficient but suffer from two critical limitations: they either avoid traffic simulation entirely, sacrificing accuracy, or apply simple rules to simplified problem representations such as road repair scheduling with only one available machine \citep{luOptimalScheduleUrban2016}.

Local search methods including Tabu Search \citep{abdzadehSimultaneousSchedulingMultiple2022} and Simulated Annealing \citep{xuStudyContinuousNetwork2009} can incorporate traffic simulation but are frequently prone to becoming trapped in local optima. The irregular objective landscape caused by complex project interactions makes it difficult for these methods to escape suboptimal regions, particularly in large-scale problems.

Population-based metaheuristics such as Genetic Algorithms \citep{cheuGeneticAlgorithmSimulationMethodology2004}, NSGA-II \citep{miralinaghiContractBundlingConsiderations2022}, Ant Colony Optimization \citep{aksoyUrbanRoadNetwork2021}, and Artificial Bee Colony \citep{jiangTimedependentTransportationNetwork2015} represent the current state-of-the-art for road work scheduling problems. These methods offer two key advantages for combinatorial scheduling problems. First, by maintaining and evolving a population of candidate solutions rather than a single solution trajectory, they preserve solution diversity throughout the search process, enabling exploration of multiple regions of the solution space simultaneously and reducing susceptibility to local optima that plague local search methods. Second, their problem-agnostic evaluation framework can accommodate complex black-box objectives where analytical gradients are unavailable \citep{katochReviewGeneticAlgorithm2021}, such as the traffic simulation-based objectives in the RNMSP-UD.

Among population-based approaches, NSGA-II is particularly well-suited for the RNMSP-UD. Its design specifically addresses multi-objective optimization through non-dominated sorting and crowding distance mechanisms, which generate diverse Pareto fronts that illuminate trade-offs between conflicting objectives. NSGA-II has been successfully applied to similar road work scheduling problems \citep{miralinaghiContractBundlingConsiderations2022, hosseininasabMultiobjectiveIntegratedModel2018}, demonstrating its effectiveness in this domain. Moreover, the modular structure of NSGA-II, where crossover and mutation operators can be customized to exploit problem-specific characteristics, provides flexibility for algorithmic enhancements. This adaptability makes NSGA-II a natural foundation upon which to build computational improvements such as our proposed method.

Beyond our problem setting, NSGA-II and related multi-objective genetic algorithms are well established in transportation infrastructure management. \citet{baiTradeOffAnalysisMultiobjective2012, baiHybridParetoFrontier2015} use non-dominated sorting to generate Pareto frontiers for trade-off analysis in infrastructure management, and \citet{ruiz-velezSustainableRoadInfrastructure2024} develop a custom NSGA-II with problem-specific repair operators for sustainable infrastructure design---mirroring our own use of tailored crossover, mutation, and repair operators. Multi-objective genetic algorithms have likewise been applied to transit network design \citep{chaiImprovedNSGAIIAlgorithm2020} and sustainable infrastructure management \citep{asadifakhrMultiObjectiveGeneticAlgorithm2025}, and, within this journal, to genetic-algorithm-based system identification of infrastructure \citep{fugginiCombiningGeneticAlgorithms2013}. These applications attest to the maturity of NSGA-II for infrastructure trade-off problems. Our contribution does not compete with them: rather than proposing another algorithm variant, we target the evaluation step---the computational bottleneck---so that such algorithms remain tractable at substantially larger problem sizes.

Population-based metaheuristics, however, face a scalability constraint in the context of expensive black-box optimization. Their effectiveness requires evaluating hundreds or thousands of candidate solutions per generation, with each evaluation in the RNMSP-UD demanding multiple traffic simulations across time periods. Furthermore, the same characteristics that enable their flexibility - independence from gradient information by relying on stochastic variation operators - also mean these methods explore the solution space through broad sampling rather than directed search towards optimal regions. As Table~\ref{table:lit_overview} demonstrates, existing applications rarely exceed 20 projects, suggesting that computational costs become prohibitive at larger scales relevant to real urban renovation programs.

Exact methods provide optimality guarantees but scale poorly to the bi-level RNMSP. Linear Programming \citep{szetoTimeDependentDiscreteNetwork2010} and Dynamic Programming \citep{maRoadMaintenanceOptimization2018} approaches require precomputing all traffic simulations beforehand, which avoids solving the coupled upper-level scheduling and lower-level assignment problems jointly but limits problem sizes to approximately 10 projects, whereas branch-and-bound, optionally combined with Benders decomposition, can address the bi-level structure directly \citep{bagloeeHybridBranchandBoundBenders2017} yet remains tractable only for small instances. 

Surrogate-assisted optimization has proven successful in other expensive simulation-optimization domains such as aerospace design and engineering optimization. Within road work scheduling, applications remain limited. Existing studies \citep{bagloeeHybridMachinelearningOptimization2018, bagloeeOptimizationRoadsConstruction2018} use linear surrogate models to approximate the performance of entire project selections or schedules—that is, they train models to predict the overall objective value of a complete solution rather than the outcome of individual traffic simulations. More recent work applies richer learned representations to road-work scheduling: \citet{saneiiMethodologySchedulingWithinday2023} combine deep neural networks with active learning to schedule within-day work zones, though still predicting at the level of a whole schedule rather than the individual traffic scenarios that compose it. Evaluating the travel-delay from a schedule is inherently more abstract, as the project start times first determine which set of roads is closed in each period of the planning horizon, and each closure set then maps, through a traffic simulation, to a congestion (travel-delay) value. A schedule-level surrogate tries to learn this entire mapping at once---from all project timings directly to the aggregate delay---whereas our surrogate learns only the second step, predicting the delay of an individual closure set. Approximating only this computationally intensive component may improve accuracy without sacrificing efficiency, and it allows exact and predicted per-period contributions to be combined within a single schedule.

More broadly, the approach we develop relates to the field of surrogate-assisted evolutionary algorithms (SAEAs), which replace or supplement expensive fitness evaluations with machine learning models. Conventional SAEAs typically train a global surrogate of the solution-level objective and integrate it into the search in standard ways, for example, pre-selecting promising offspring or replacing expensive fitness evaluations with the surrogate's prediction. Our method differs in three respects that are central to its novelty. First, it approximates at the level of an individual traffic scenario rather than the aggregate schedule objective, so that exact and estimated contributions can be combined within a single solution. Second, the surrogate is used not to replace fitness but to construct a conservative lower-bound estimate that drives a dominance-based pruning rule: a candidate is discarded once its partly exact, partly estimated objective appears dominated. This resembles the bounding step of branch-and-bound, though the estimate is conservative rather than a provable bound, so a pruned candidate is almost always genuinely dominated, and the rare incorrect prune has a negligible effect on the final Pareto front rather than compromising it. Third, this estimate is refined progressively, one scenario at a time, which keeps pruning conservative and empirically preserves solution quality. To our knowledge, this combination of simulation-level surrogacy, conservative lower-bound estimation, and progressive refinement for pruning within a multi-objective evolutionary algorithm has not previously been applied to road work scheduling.

\subsection{Identified Gaps and Research Contribution}
\label{subsec: literature gap}
Our literature analysis reveals three limitations in existing road work scheduling research that motivate this study's contributions.

First, existing research typically treats infrastructure lifespans as deterministic constraints with fixed deadlines. To our knowledge, no prior study incorporates the probabilistic risk of failure as a primary objective. By contrast, our model treats deadlines as "soft" by explicitly accounting for the expected cost of failure—defined as the probability of collapse multiplied by its associated impact—alongside traffic disruption. This allows for a more nuanced optimization of the trade-off between infrastructure longevity and user convenience.

Second, computational scalability remains constrained. Table \ref{table:lit_overview} shows that most studies incorporating traffic simulation address fewer than 35 projects, with the largest handling 100 projects without traffic simulation \citep{yamanyNetworklevelPavementMaintenance2024}. This raises questions about applicability to real urban networks requiring coordination of dozens or hundreds of projects. This challenge is particularly acute for multi-objective formulations, which inherently require finding diverse solution sets rather than single optimum. 

Third, existing applications of surrogate modeling to road work scheduling appear limited in scope. Studies that employ surrogate models \citep{bagloeeHybridMachinelearningOptimization2018, bagloeeOptimizationRoadsConstruction2018} train their models to predict the overall objective value of an entire schedule, yet this approach foregoes the granularity needed for progressive evaluation. A schedule's total travel delay aggregates many individual traffic simulations—one for each combination of active projects across time periods. By approximating at the simulation level rather than the schedule level, surrogates can enable selective evaluation: computing some time periods exactly while estimating others, and refining assessments incrementally. To our knowledge no research has explored using surrogates to eliminate dominated solutions during optimization, despite the potential for such strategies to reduce computational requirements substantially.

This paper addresses these limitations through three corresponding contributions. First, we formulate the RNMSP-UD as a bi-level multi-objective optimization problem that explicitly models uncertain infrastructure lifespans through probabilistic failure models, introducing failure risk as an objective alongside traffic disruption. Second, we develop Progressive Lower Bound Evaluation (PLBE), which integrates machine learning surrogate models with NSGA-II to reduce required traffic simulations. PLBE uses surrogates to progressively estimate solution performance and eliminate dominated candidates early, targeting the specific bottleneck rather than overall solution quality. Third, we demonstrate PLBE's effectiveness on the Sioux Falls network with 76 projects—substantially larger than typical scales in existing literature—achieving up to an 87$\times$ increase in completed optimization iterations within the same time budget (and substantially fewer traffic simulations) compared to standard NSGA-II while improving solution quality across multiple performance metrics.

\section{Problem Formulation}
\label{sec:problem}

In this section, we formally define the Road Network Maintenance Scheduling Problem with Uncertain Deadlines (RNMSP-UD) as a bi-level optimization problem. The formulation consists of an upper level scheduling problem controlled by the decision maker and a lower level traffic assignment problem that models road users' routing responses to infrastructure closures. We begin by establishing the general problem context before presenting the mathematical formulation of each level in Sections~\ref{subsec:ULproblem} and~\ref{subsec:LLproblem}, respectively.

\begin{sidewaystable*}[p]
\centering
\caption{Overview of reviewed literature.}
\label{table:lit_overview}
\small
\setlength{\tabcolsep}{4pt}
\begin{tabular}{l l ll l l l l l l}
\toprule
\textbf{Paper} & \textbf{Problem Class} &\textbf{Scheduling}& \textbf{Objectives} & \textbf{Solution Method} & \textbf{MO}& \textbf{Case Study} & \textbf{Projects} & \textbf{Uncert.} & \textbf{Surrogate} \\

\citet{changTabuSearchBased} &  WZSP&\checkmark& TTD& Tabu Search & & Columbus & 6& -- & --\\
\citet{cheuGeneticAlgorithmSimulationMethodology2004} &  RNMSP&\checkmark& TTD & GA & & Clementi & 10& -- & --\\
\citet{poorzahedyApplicationAntSystem2005} &  NDP&& TTD & ACO& & Sioux Falls & 10& -- & --\\
\citet{taoIslandModelsStochastic2007} &  NDP&\checkmark& TTD & Island Model & & Synthetic & 20& -- & --\\
\citet{leeOptimizingScheduleImproving2009} &  RNMSP&\checkmark& TTD & ACO& & Taitung City & 33& -- & --\\
\citet{ukkusuriMultiperiodTransportationNetwork2009} &  NDP&\checkmark& CS&   SQP& &   Nguyen Dupius&   12& -- & --\\
\citet{szetoTimeDependentDiscreteNetwork2010} &  NDP&\checkmark& Social Surplus & LP & & Synthetic & 4& -- & --\\
\citet{chienSchedulingHighwayWork2014} &  WZSP&\checkmark& TTD, Cost & GA & & Middlesex & 5& -- & --\\
\citet{jiangTimedependentTransportationNetwork2015} &  NDP&\checkmark& CS, Health Cost& ABC& & Sioux Falls & 8& -- & --\\
\citet{gongOptimizingSchedulingLongterm2016} &  RNMSP&\checkmark& TTD & GA & & Sioux Falls & 5& -- & --\\
\citet{luOptimalScheduleUrban2016} &  RNMSP&\checkmark& TTD & Greedy Algorithm & & Sioux Falls & 12& -- & --\\
\citet{bagloeeOptimizationRoadsConstruction2018}&  NDP&\checkmark& Cost, TTD & ML + LP& \checkmark& Winnipeg & 20& -- & Schedule\\
\citet{kumarSimplifiedFrameworkSequencing2018} &  NDP&\checkmark& User Benefit & Cost-Benefit Heuristic & & Montgomery & 40& -- & --\\
\citet{zhengMeasuringNetworkwideTraffic2014} &  RNMSP&\checkmark& TTD & Local Search & & Guam & 6& -- & --\\
\citet{maRoadMaintenanceOptimization2018} &  RNMSP&\checkmark& TTD, Cost & DP & \checkmark& Sioux Falls & 10& -- & --\\
\citet{songRehabilitationProjectSelection2018} &  RNMSP&\checkmark& NPV & Active Set Algorithm & & Sioux Falls & 10& -- & --\\
\citet{hosseininasabMultiobjectiveIntegratedModel2018} &  NDP&\checkmark& TTD, Equity, SS & Enhanced NSGA-II & \checkmark& Sioux Falls & 10& -- & --\\
\citet{miralinaghiNetworklevelSchedulingRoad2020} &  NDP&\checkmark& Total Costs& Active Set Algorithm & & Sioux Falls & 18& -- & --\\
\citet{miralinaghiContractBundlingConsiderations2022} &  NDP&\checkmark& TTD, Cost & NSGA-II & \checkmark& Anaheim & 12& -- & --\\
\citet{liBilevelOptimizationLongterm2021} &  RNMSP&\checkmark& TTD & GA & & Sioux Falls & 5& -- & --\\
\citet{aksoyUrbanRoadNetwork2021} &  RNMSP&\checkmark& TTD & ACO& & Synthetic & 12& -- & --\\
 \citet{moghtadernejadPrioritizingRoadNetwork2022} & RNMSP& \checkmark& Cost, TTD& PSO& & Chur&50& -- & --\\
 \citet{seilabiReinforcementLearningbasedApproach2025} & RNMSP& \checkmark& Total Emissions& RL + PSO& & Anaheim &6& -- & --\\
 \citet{bagloeeHybridMachinelearningOptimization2018} & NDP& & TTD& ML + LP& & Winnipeg &20& -- & Schedule\\
\citet{yamanyNetworklevelPavementMaintenance2024}& NDP& \checkmark& TTD& GA & & Wyoming&100& -- & --\\
\textbf{This paper}&    \textbf{RNMSP}&\checkmark&   \textbf{TTD, Risk}&   \textbf{ML + NSGA-II}&   \checkmark&   \textbf{Sioux Falls }&   \textbf{76}& \textbf{\checkmark} & \textbf{Simulation}\\
\bottomrule
\end{tabular}

{\footnotesize\textit{Notes:} Scheduling = inclusion of time in the problem definition; MO = multi-objective solution method; Uncert.\ = models uncertain deadlines/lifespans; Surrogate = level of surrogate approximation (none, schedule-level, or simulation-level). Abbreviations: ACO, Ant Colony Optimization; SQP, Sequential Quadratic Programming; ABC, Artificial Bee Colony; LP, Linear Programming; PSO, Particle Swarm Optimization; RL, Reinforcement Learning. The Yamany et~al.\ study addresses project sequencing rather than scheduling per se.\par}
\end{sidewaystable*}

\subsection{Problem Structure}

We consider a set of infrastructure elements (roads, bridges, quay walls, or sewage systems) requiring renovation, where each renovation obstructs access to one or more road segments. The decision maker—typically a municipality or infrastructure authority—determines the starting time for each project within a finite planning horizon divided into discrete time periods (e.g., quarterly intervals over ten years). These represent strategic planning decisions---determining which projects begin in which period across a multi-year horizon---as opposed to operational work-zone decisions, such as work hours or lane-closure configurations on individual roads, which belong to a separate downstream problem \citep{changTabuSearchBased, chienSchedulingHighwayWork2014}. Infrastructure lifespans are inherently uncertain, making fixed deadlines unsuitable. We therefore model this uncertainty through soft deadlines representing the risk of emergency intervention—computed as the cumulative probability that infrastructure reaches a critical state before the scheduled start time multiplied by emergency intervention costs—as formalized in Section~\ref{subsec:ULproblem}.

Scheduling decisions are constrained by available budget and limits on simultaneous project execution, reflecting real-world limitations on labor, materials, and contractor capacity. Traffic flow is simulated at each time period to assess renovation impacts on local traffic. Road closures affect route choices and generally increase total travel time. The decision maker balances two conflicting objectives: minimizing expected emergency intervention cost (Risk $R$) and minimizing the increase in congestion (Total Travel Delay $TTD$) relative to a baseline without disruptions.

The RNMSP-UD exhibits a bi-level structure arising from the interaction between centralized scheduling and decentralized routing. At the upper level, the decision maker selects a schedule minimizing both failure risk and traffic disruption, subject to budgetary and resource constraints. At the lower level, road users independently choose routes minimizing their personal travel time. Table~\ref{tab:bilevel_structure} summarizes this structure.

\begin{table*}[h]
\centering
\begin{tabularx}{\textwidth}{X X X X X}   
\toprule
 & Problem & Decisions & Decision-maker & Objective\\
\midrule
Upper level & Scheduling & Project starting times & Municipality & Minimize $R$ \& $TTD$\\
Lower level & Traffic Assignment & Route choices & Road users & Minimize individual travel time\\
\bottomrule
\end{tabularx}
\caption{Structure of the bi-level RNMSP-UD formulation}
\label{tab:bilevel_structure}
\end{table*}

This structure creates sequential dependency: scheduling choices directly affect road capacity near ongoing renovation projects, which affects routing decisions of drivers at the lower level. Thus, schedule performance can only be determined by evaluating aggregate routing decisions from the lower level problem. Each candidate schedule requires traffic simulations across all time periods in the planning horizon to compute  $TTD$, where these simulations solve the lower level problem. This bi-level structure introduces significant computational challenges through repeated expensive traffic simulations. The following sections formalize each problem level.

\subsection{Upper Level Problem}
\label{subsec:ULproblem}

In this section, we formalize the upper level scheduling problem, which determines the starting times of renovation projects to minimize both infrastructure failure risk and traffic disruption. We present the objectives, decision variables, and constraints that govern feasible scheduling decisions. The upper level problem is a static multi-objective scheduling problem in which the decision maker determines the start times of infrastructure renovation projects, subject to constraints on budgets, deadlines, and the maximum number of simultaneous construction activities. The complete notation is provided in~\ref{table:UL_notation}.

The decision maker minimizes two conflicting objectives; the total risk $R$ and the total travel delay $TTD$. The risk objective, which represents the expected cost of emergency interventions incurred when infrastructure deteriorates before its scheduled renovation, forms the primary novelty of our problem formulation compared to existing literature on road work scheduling. Risk is defined as the chance of emergency intervention multiplied by the cost per project:

\begin{equation}\label{eq:risk}
    R = \sum_{p \in \mathcal{P}} \sum_{t \in \mathcal{T}} f_{tp} X_{tp} w_p
\end{equation}

Here, $f_{tp}$ represents the probability that infrastructure $p$ reaches a critical state before the renovation begins in period $t$, modeled as the survival function of a binomial distribution---the probability that the number of deterioration events, each occurring with per-period probability $v_p$, exceeds the project's tolerance $k_p$:

\begin{equation}
    {
    f_{tp} = 1 - F(k_p, t, v_p) = 1 - \sum_{i=0}^{k_p} \binom{t}{i} (v_p)^i (1 - v_p)^{t - i}}
    \label{eq:cdf_binomial}
\end{equation}

We model deterioration as a discrete-time hazard process: in each elapsed period the asset independently experiences a deterioration event with probability $v_p$, and emergency intervention becomes necessary once the cumulative number of such events exceeds a project-specific tolerance $k_p$. The resulting survival function is the discrete-time counterpart of the hazard-based deterioration models widely used in modelling infrastructure reliability, where the probability of failure grows monotonically with age under uncertain, nonlinear degradation~\citep{biondiniLifeCyclePerformanceDeteriorating2016}. The expected risk function exhibits sigmoid-shaped growth: the probability $f_{tp}$ asymptotically approaches 1, meaning that the risk is bounded above by $w_p$. Importantly, the marginal increase in expected risk per delay period is greatest when $f_{tp} = 0.5$ and then declines. Without intervention, this declining gradient could perversely incentivize delaying projects indefinitely as they approach near-certain failure, which is neither desirable nor realistic. To prevent such strategies, we impose a hard deadline at the point where $f_{tp} > 0.5$, ensuring that the optimization promotes proactive rather than reactive scheduling. This threshold has a natural interpretation: since $f_{tp}$ is the cumulative failure probability, the deadline falls in the period by which failure has become more likely than not---the asset's median lifetime. It also lies near the inflection point of the sigmoid risk curve, beyond which the declining marginal risk would begin to reward further postponement. Deadlines of this form follow the control-limit policies common in condition-based maintenance, where intervention is triggered once a condition or failure-risk measure crosses a prescribed level~\citep{zhangOptimalConditionbasedMaintenance2024}. We acknowledge that the specific value remains a modeling choice: an agency could adopt a more conservative threshold than $0.5$, optimize over simulated lifespan realizations in a stochastic-programming formulation, or treat intervention timing as a sequential decision problem (a Markov decision process). We deliberately retain the fixed median-lifetime deadline because it accounts for uncertain lifespans while avoiding the additional modeling and computational complexity that stochastic dynamic programming or reinforcement-learning formulations would introduce; we return to these extensions in Section~\ref{sec:conclusion}. The binomial form is a modeling convenience rather than a restriction: nothing in the optimization method depends on it. Any monotonically non-decreasing cumulative failure probability $f_{tp}$---for example a Weibull lifetime distribution or a Markovian condition-state process~\citep{zhangOptimalConditionbasedMaintenance2024}---can be substituted, and the hard-deadline mechanism at $f_{tp} = 0.5$ carries over unchanged.

The second objective, total travel delay $TTD$, quantifies the aggregate increase in travel time across the network due to ongoing renovation projects. This metric captures the societal cost of traffic disruption over the planning horizon:

\begin{equation}\label{eq:ttd}
    {TTD = \sum_{t \in \mathcal{T}} \sum_{(i,j) \in \mathcal{A}} f_{ij}^t \cdot c_{ij}^t - |\mathcal{T}| \cdot tt_{\text{base}}}
\end{equation}

Here, $f_{ij}^t$ and $c_{ij}^t$ denote the equilibrium traffic flow and travel time on arc $(i,j)$ during period $t$ under the network conditions created by ongoing renovations, so that the inner sum is the flow-weighted total system travel time of period $t$ (in vehicle-hours), and $tt_{\text{base}} = \sum_{(i,j)\in\mathcal{A}} f_{ij}^{0}\, c_{ij}^{0}$ is the same quantity in the undisturbed network. The equilibrium flows and travel times are not known \textit{a priori} but must be determined by solving the lower level traffic assignment problem for each time period given the set of active projects (as formalized in Section~\ref{subsec:LLproblem}). Thus, $TTD$ measures the cumulative excess travel time experienced by all road users throughout the planning horizon due to the capacity reductions imposed by the renovation schedule.

The upper level problem is subject to several constraints that ensure feasible scheduling decisions. These constraints enforce that each project starts exactly once, projects cannot start too late to finish before their hard deadlines, ongoing projects continue until completion, the number of simultaneous projects is limited, and budget constraints are satisfied at all time periods. The complete mathematical formulation of the upper level problem is provided in \ref{appendix:UL_formulation}.

This formulation of the upper level scheduling problem establishes the decision-making framework for minimizing emergency intervention costs while mitigating traffic disruption. The risk objective $R$ encourages timely interventions under uncertain infrastructure lifespans, while the $TTD$ objective internalizes the effects of renovations on road users. In Section~\ref{subsec:LLproblem}, we formalize the lower level problem, which models how traffic flows respond to scheduled renovations.

\subsection{Lower Level Problem}
\label{subsec:LLproblem}

In this section, we formalize the lower level traffic assignment problem, which determines how traffic flows distribute across the road network in response to capacity reductions caused by ongoing renovation projects. This allocation enables computation of the total travel delay objective defined in the upper level problem.

The lower level problem allocates traffic flows over the traffic network $\mathcal{G}=(\mathcal{N},\mathcal{A})$ for each discrete time period $t \in \mathcal{T}$ within the planning horizon. This allocation enables us to compute the total travel delay ($TTD$) in Equation~\ref{eq:ttd}. Traffic demand consists of a predefined set of trips between origin-destination (O-D) pairs in $\mathcal{G}$. These trips are specified as peak-hour volumes, representing the most congested period of a typical day. To calculate the impact over the planning horizon, the simulation results from this peak period are treated as a representative daily load and are subsequently extrapolated to match the temporal granularity of the horizon, such as monthly or yearly totals. 

The assignment of traffic follows the principle of User Equilibrium, which states that no driver can change routes unilaterally in order to reduce their travel time. This problem, known as the Traffic Assignment Problem (TAP), is a convex non-linear mathematical program~\citep{nguyenAlgorithmTrafficAssignment1974}. Mathematically, the equilibrium flow pattern solves a single convex program---the Beckmann transformation \citep{beckmannStudiesEconomicsTransportation1956}---whose objective integrates each link's travel-cost function over the total flow on that link, as shown in Equation~\ref{eq:UE_objective}. Its minimizer satisfies Wardrop's First Principle, expressed by the complementarity conditions below, under deterministic flow assumptions; note that this objective is a mathematical device whose minimizer characterizes the equilibrium distribution, not a travel cost that is being minimized. The results of the TAP change in reaction to the temporary removal or capacity reduction of roads where a project is ongoing. The complete notation for the traffic network model is provided in \ref{appendix:LL_formulation}.

\begin{equation}
\label{eq:UE_objective}
    {\min_{\{x_\pi\}} \sum_{(i,j) \in \mathcal{A}} \int_0^{f_{ij}} c_{ij}(z) \, dz, \qquad f_{ij} = \sum_{w \in \mathcal{W}} \sum_{\pi \in \Pi_w} \delta_{ij}^\pi \, x_\pi}
\end{equation}
\[
{\text{subject to:} \quad \sum_{\pi \in \Pi_w} x_\pi = m_w, \quad x_\pi \geq 0 \quad \forall w \in \mathcal{W}}
\]
\[
{c_\pi - \lambda_w \geq 0, \quad (c_\pi - \lambda_w) x_\pi = 0 \quad \forall \pi \in \Pi_w}
\]
where $c_{ij}(\cdot)$ is the BPR link-cost function evaluated at the total link flow and $\delta_{ij}^\pi$ is the link--path incidence indicator (notation as in \ref{appendix:LL_formulation}).

The formulation is subject to several constraints that ensure realistic traffic flow distribution. These constraints enforce non-negative path flows, link flow conservation (where link flows equal the sum of path flows using that link), path cost calculations based on link costs, congestion modeling through the Bureau of Public Roads function, and dynamic capacity adjustments when renovation projects are ongoing. The complete mathematical notation for the traffic network model is provided in Table~\ref{table:LL_notation}, while the detailed algorithmic formulation is presented in \ref{appendix:LL_formulation}.

The TAP ensures that traffic flows are allocated according to realistic user behavior while incorporating the effects of road renovation projects. The problem is formulated as a convex nonlinear program, where user equilibrium ensures a self-regulating flow distribution. The constraints ensure demand fulfillment, flow conservation, and dynamic capacity adjustments due to ongoing projects, while the congestion function models traffic delay effects. This formulation allows us to compute the total travel delay $TTD$, which is essential to evaluate the impact of renovation schedules at the upper level.

\section{Solution Method}
\label{sec:solmethod}

As outlined in Section~\ref{sec:problem}, the problem is formulated as a bi-level optimization problem, consisting of a scheduling problem at the upper level and a Traffic Assignment Problem (TAP) at the lower level. This section presents the solution methodologies employed for both levels. For the upper level problem, we utilize the Non-Dominated Sorting Genetic Algorithm II (NSGA-II), which is a Genetic Algorithm (GA) tailored specifically for multi-objective optimization \citep{debFastElitistMultiobjective2002, miralinaghiContractBundlingConsiderations2022, hosseininasabMultiobjectiveIntegratedModel2018}. To enhance its efficiency, we introduce Progressive Lower Bound Evaluation (PLBE). Both methods are discussed in Section~\ref{subsec:ulsm}. For solving the lower level problem, the Frank-Wolfe algorithm is employed to allocate traffic in accordance with User Equilibrium conditions. A detailed explanation of this algorithm is given in Section~\ref{subsec:llsm}.

\subsection{Solution Method Architecture}
\label{subsec:solmethod_architecture}
\begin{figure*}[htpb]
    \centering
    \includegraphics[width=0.5\linewidth]{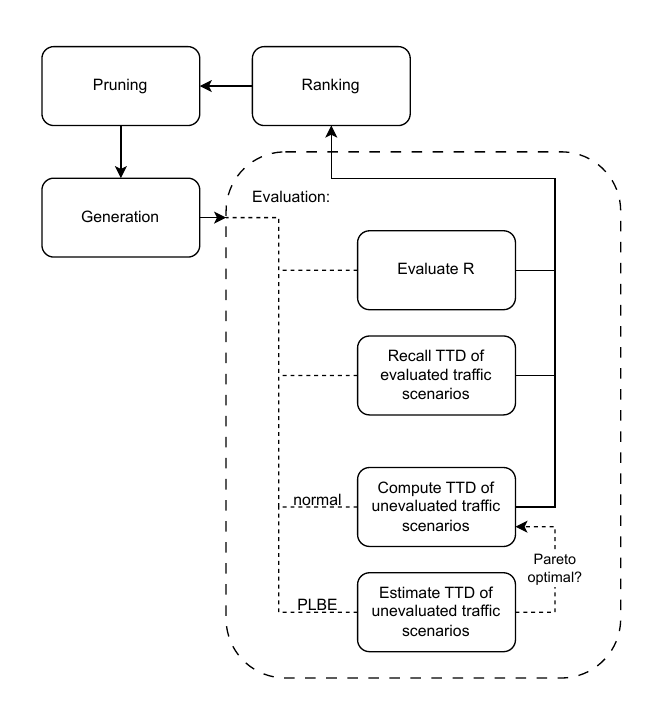}
    \caption{Integration of Progressive Lower Bound Evaluation (PLBE) into the NSGA-II framework. The standard evaluation (top path) computes TTD for all traffic scenarios using Frank-Wolfe. PLBE (bottom path) first estimates TTD for unevaluated scenarios; full computation proceeds only for solutions estimated to be Pareto optimal.}
    \label{fig:solution_architecture}
\end{figure*}

This section provides an overview of our solution approach before detailing individual components. The RNMSP-UD is solved through a hierarchical architecture where NSGA-II addresses the upper level scheduling problem while the Frank-Wolfe algorithm solves the lower level TAP. Figure~\ref{fig:solution_architecture} illustrates how PLBE integrates into the NSGA-II framework.

NSGA-II operates through iterative cycles consisting of four phases: generation of offspring solutions through crossover and mutation operators, evaluation of solution performance, ranking using non-dominated sorting, and pruning to select the best individuals for the next generation. The evaluation phase, shown in Figure~\ref{fig:solution_architecture}, comprises two objectives with markedly different computational requirements. The risk objective $R$ follows directly from the probabilistic failure model and is computationally inexpensive. In contrast, the traffic disruption objective $TTD$ requires solving multiple Traffic Assignment Problems—one for each time period in the planning horizon—using the Frank-Wolfe algorithm to determine how traffic flows redistribute under different road closure configurations.

Our methodological contribution modifies the evaluation phase through PLBE, shown in the lower path of Figure~\ref{fig:solution_architecture}. Rather than immediately computing $TTD$ for all traffic scenarios using Frank-Wolfe, PLBE first estimates performance for unevaluated traffic scenarios using machine learning surrogate models. Each candidate schedule spans multiple time periods, each with a traffic scenario defined by the active projects during that period. Since some time periods have identical closures, the set of unique traffic scenarios $\mathcal{S}$ is a subset of all traffic scenarios, with each unique scenario requiring the outcome of a traffic simulation. Consider $\mathcal{S}_{eval} \subseteq \mathcal{S}$ as the set of scenarios with known simulation outcomes and $\mathcal{S}_{remaining} = \mathcal{S} \setminus \mathcal{S}_{eval}$ as the set of scenarios yet to be evaluated. For scenarios belonging to $\mathcal{S}_{eval}$, exact $TTD$ values are recalled from memory. For those in $\mathcal{S}_{remaining}$, the surrogate model provides estimates. If this combination of exact and estimated values indicates the solution is dominated by existing candidates, evaluation terminates. Otherwise, one scenario from $\mathcal{S}_{remaining}$ is simulated using Frank-Wolfe, the estimate is refined, and dominance is reassessed. This iterative process continues until the solution is either discarded as dominated or fully evaluated. By concentrating computational effort on promising candidates, PLBE achieves substantial efficiency gains; the small quality loss from rare pruning errors (quantified in Section~\ref{subsec:exp1_new}) is more than compensated by the additional iterations these gains enable, with a significant net improvement in solution quality across the entire front (Section~\ref{sec:experiments}).

\subsection{Upper Level Solution Method}
\label{subsec:ulsm}

In this section, we present our solution approach for the upper level scheduling problem. As established in Section~\ref{sec:literature}, we employ NSGA-II as our base optimization algorithm due to its suitability for multi-objective problems and its flexibility in handling optimization problems where gradients are not available. To enhance NSGA-II's computational efficiency in optimizing expensive black-box simulation, we introduce Progressive Lower Bound Evaluation (PLBE), a method that aims to reduce the number of traffic simulations required. Section~\ref{subsubsec:NSGA} describes the NSGA-II framework, and Section~\ref{subsubsec:PLBE} introduces the PLBE method.

\subsubsection{NSGA-II}
\label{subsubsec:NSGA}

NSGA-II is a multi-objective evolutionary algorithm designed to optimize problems involving multiple conflicting objectives. Unlike conventional GAs, NSGA-II generates a partial Pareto front of non-dominated solutions, facilitating a trade-off analysis among competing objectives. The algorithm uses natural selection principles to iteratively refine a population of candidate solutions.

Each solution is represented as a genotype, a structured data string encoding the solution parameters. In the context of this study, the genotype consists of an integer sequence that specifies the start time for all renovation projects under consideration. During each iteration, the fitness of individuals in the population is assessed based on the objective functions, and the best-performing candidates are selected. Subsequently, new offspring are generated using crossover and mutation operators. The crossover operation combines portions of the integer sequences from two parent individuals to create a new candidate solution, while the mutation operator introduces small perturbations to project start times to diversify the search space. The newly generated offspring are evaluated, and the entire population is ranked using non-dominated sorting~\citep{debFastElitistMultiobjective2002}. The top-performing individuals are retained based on their Pareto rank and, in cases of ties, crowding distance. This process is repeated until the termination criteria are satisfied. The pseudo-code for NSGA-II is outlined in Algorithm~\ref{algo:NSGA}.

\begin{algorithm}
\caption{Non-Dominated Sorting Genetic Algorithm-II}
\label{algo:NSGA}
\begin{algorithmic}[1]
    \State Initialize the population with randomly generated individuals.
    \State Evaluate the fitness of each individual.
    \While {termination criteria are not met}
        \State Select parent solutions using tournament selection.
        \State Apply crossover and mutation operators to generate offspring.
        \State Evaluate the fitness of the offspring.
        \State Merge parent and offspring populations.
        \State Perform non-dominated sorting to assign Pareto front ranks.
        \State Select the top individuals using rank and crowding distance.
        \State Update the population.
    \EndWhile
    \State \Return the Pareto front of non-dominated solutions.
\end{algorithmic}
\end{algorithm}

\subsubsection{Progressive Lower Bound Evaluation}
\label{subsubsec:PLBE}

This section introduces Progressive Lower Bound Evaluation (PLBE), a method that builds upon our preliminary work on surrogate modeling for traffic prediction~\citep{boschMachineLearningPredictions2025} to enhance the computational efficiency of NSGA-II when applied to the RNMSP-UD. While NSGA-II is well-suited for handling the complex, multi-objective and combinatorial nature of the problem, its scalability is constrained by the cost of evaluating individual solutions. In our setting, each solution requires the simulation of traffic conditions over a multi-period network with varying sets of road closures. Because the number of required simulations grows exponentially with the number of projects under consideration, a single generation of NSGA-II may demand tens of thousands of simulations, making evaluation the primary computational bottleneck.

PLBE addresses this bottleneck by reducing the number of full evaluations through an iterative estimation strategy. Each candidate solution specifies which projects are ongoing in each time period of the planning horizon, thereby determining which roads are blocked at each point in time. Congestion in any given time period depends on the specific combination of blocked roads during that period. We refer to each unique set of blocked roads in the network as a traffic scenario. Computing the total travel delay ($TTD$) for a solution requires determining the congestion outcome associated with each traffic scenario that occurs throughout the planning horizon.

The core mechanism of PLBE operates as follows. For each candidate solution with traffic scenarios $\mathcal{S} = \{s_1, s_2, \ldots, s_n\}$, we maintain two sets: $\mathcal{S}_{eval} \subseteq \mathcal{S}$ denoting scenarios with known simulation outcomes, and $\mathcal{S}_{remaining} = \mathcal{S} \setminus \mathcal{S}_{eval}$ denoting scenarios yet to be evaluated. The method begins by computing the risk objective $R$ for each solution (which is computationally inexpensive) and providing a conservative estimate of the total travel delay ($TTD$) by combining simulated results from $\mathcal{S}_{eval}$ with surrogate model predictions for scenarios in $\mathcal{S}_{remaining}$. If this initial estimate suggests the solution could be Pareto optimal, the method evaluates the most frequently occurring scenario from $\mathcal{S}_{remaining}$—moving it to $\mathcal{S}_{eval}$—and updates the $TTD$ estimate. This iterative process continues, alternating between checking for dominance and simulating one additional scenario, until either $\mathcal{S}_{remaining} = \emptyset$ (full evaluation completed) or the updated estimate indicates the solution is dominated and can be discarded. This form of progressive evaluation is especially effective in later generations of the algorithm, where the quality of Pareto-optimal solutions is high and the accuracy of the surrogate estimates increases due to the availability of more simulation data.

We implement two variants of PLBE that differ in how they handle solutions identified as dominated before full evaluation. In one variant (that we name Elimination Pruning), when a solution is found to be dominated based on estimated performance $(\widehat{TTD}, R)$, it is permanently marked with this estimate and excluded from future generations. This approach risks incorrectly eliminating promising solutions due to surrogate underestimation, but avoids redundant evaluations of suboptimal solutions across generations. In the other variant (that we name Lazy Evaluation), dominated solutions are simply skipped without permanent marking, allowing them to be regenerated and re-evaluated in future generations when surrogate accuracy may have improved. This reduces the risk of permanently losing good solutions but may result in redundant evaluations. Algorithm~\ref{algo:PLBE} presents the unified procedure, with lines 12--14 implementing the variant-specific handling: Elimination Pruning permanently marks pruned solutions while Lazy Evaluation does not. The choice between variants thus represents a tradeoff between computational efficiency and robustness to surrogate error---Elimination Pruning achieves higher throughput when surrogates are accurate, while Lazy Evaluation provides a safeguard when surrogate quality is uncertain.

\begin{algorithm}[htbp]
\caption{Progressive Lower Bound Evaluation}
\label{algo:PLBE}
\begin{algorithmic}[1]
\Require Population $P$, offspring population $Q$, scenario set $\mathcal{S}$, surrogate model $f_{surrogate}$, variant $\in \{\text{EliminationPruning}, \text{LazyEvaluation}\}$
\Ensure Evaluated population $Q'$

\State $Q' \leftarrow \emptyset$
\For{each individual $x \in Q$}
    \State $\mathcal{S}_{eval}(x) \leftarrow \emptyset$
    \State $\mathcal{S}_{remaining}(x) \leftarrow \mathcal{S}$
    \State $R(x) \leftarrow$ ComputeRisk$(x)$
    \State $pruned \leftarrow$ \textbf{false}
    
    \While{$\mathcal{S}_{remaining}(x) \neq \emptyset$ \textbf{and not} $pruned$}
        \State $\widehat{TTD}(x) \leftarrow$ EstimateTTD$(x, \mathcal{S}_{eval}(x),$
        \Statex \hspace{\algorithmicindent}\hspace{\algorithmicindent}\hspace{\algorithmicindent}$\mathcal{S}_{remaining}(x), f_{surrogate})$
        
        \If{IsDominated$(x, P \cup Q', (\widehat{TTD}(x), R(x)))$}
            \State $pruned \leftarrow$ \textbf{true}
            \If{variant $=$ EliminationPruning}
                \State Mark $x$ with performance $(\widehat{TTD}(x), R(x))$ \Comment{Permanent}
            \EndIf
        \Else
            \State $s^* \leftarrow$ SelectScenario$(\mathcal{S}_{remaining}(x))$
            \State $TTD_{s^*}(x) \leftarrow$ SimulateTraffic$(x, s^*)$
            \State $\mathcal{S}_{eval}(x) \leftarrow \mathcal{S}_{eval}(x) \cup \{s^*\}$
            \State $\mathcal{S}_{remaining}(x) \leftarrow \mathcal{S}_{remaining}(x) \setminus \{s^*\}$
        \EndIf
    \EndWhile
    
    \If{\textbf{not} $pruned$ \textbf{and} $\mathcal{S}_{remaining}(x) = \emptyset$}
        \State $TTD(x) \leftarrow$ ComputeExactTTD$(\mathcal{S}_{eval}(x))$
        \State $Q' \leftarrow Q' \cup \{x\}$
    \EndIf
\EndFor

\State \Return $Q'$
\end{algorithmic}
\end{algorithm}

A candidate $x$ with objective estimate $(\widehat{TTD}(x), R(x))$ is compared against the current non-dominated (incumbent) set under standard Pareto dominance: $x$ is dominated if some incumbent solution is no worse in both objectives and strictly better in at least one. The estimate $\widehat{TTD}(x)$ is a conservative estimate rather than a guaranteed lower bound: the surrogate is trained to under-predict (see below) and the heuristic returns the largest delay among evaluated sub-scenarios, so $\widehat{TTD}(x)$ is unlikely to substantially overestimate the true delay, though Braess's paradox precludes a formal guarantee. Consequently, discarding a candidate that appears dominated even under its conservative estimate rarely removes a truly non-dominated schedule, though it may occasionally do so when the surrogate over-predicts. This residual risk is exactly what separates the two variants: Lazy Evaluation discards such candidates for the current generation only, allowing them to reappear later, whereas Elimination Pruning records the estimate to suppress regeneration, trading robustness for throughput.

Drawing on our prior work on surrogate modeling for traffic prediction~\citep{boschMachineLearningPredictions2025}, we implement two surrogate models for estimating $TTD$: the CostliestSubsetHeuristic and XGBoost trained with a Pinball loss function. The CostliestSubsetHeuristic is described in~\ref{appendix:heuristic}. XGBoost was selected as the machine learning surrogate after comparative evaluation of multiple regression models~\citep{boschMachineLearningPredictions2025}, demonstrating superior performance for this application. Its tree-based structure is particularly well-suited for handling the high-dimensional binary input data that characterizes our problem where each input consists of binary indicators specifying whether project $p$ is ongoing in traffic scenario $s_i$ as well as the exponentially distributed output values typical of traffic congestion in networks~\citep{chenXGBoostScalableTree2016}. Through this combination of progressive evaluation and surrogate modeling, PLBE enables the optimization of substantially larger problem instances than would be feasible with standard NSGA-II.

The XGBoost surrogate is a gradient-boosted regression-tree ensemble (1000 trees, maximum depth 8, leaf-wise growth, learning rate 0.1, $L_2$ regularization 0.1, histogram-based split finding) trained with the quantile (pinball) loss at quantile $q$. Its inputs are sparse binary vectors indicating which projects are under renovation in a given traffic scenario, and its targets are the exact total travel delays of the scenarios simulated so far during the search. The model is retrained online whenever at least 1000 new scenarios have been simulated, using a 90/10 training/validation split with early stopping (patience 10). To keep estimates conservative, the surrogate is trained at low quantiles so that it tends to under- rather than over-predict, and its predictions are floored at the empty-network travel delay, the minimum attainable value. Training therefore requires no offline data collection: the surrogate bootstraps from simulations the algorithm performs anyway, and its accuracy improves as the search proceeds. On this task the XGBoost surrogate achieved the lowest pinball loss among the regression models evaluated in our companion study~\citep{boschMachineLearningPredictions2025}, with a mean absolute percentage error that falls from approximately 16\% after 10{,}000 simulations to 11\% after 200{,}000 as the training set accumulates. The surrogate's own computational overhead is minor: for the experimental configurations of Section~\ref{sec:experiments} (49{,}000--122{,}000 simulations accumulated over 24 hours, hence 49--122 online retrainings), a single-core benchmark of this exact configuration puts the cumulative cost of all retrainings below $1\%$ of the computational budget, and a batch prediction costs on the order of $10\,\mu\text{s}$ per scenario. The computational bottleneck thus remains the traffic assignment itself.

\subsection{Lower level Solution Method}
\label{subsec:llsm}

The lower level problem corresponds to the Traffic Assignment Problem with User Equilibrium (UE-TAP), in which no traveler can reduce their travel time by unilaterally altering their route choices. To solve this problem, we employ the Frank-Wolfe algorithm, which is the most widely used method for calculating user equilibrium solutions in transportation networks~\citep{nguyenAlgorithmTrafficAssignment1974}. 

The Frank-Wolfe algorithm is an iterative descent method that exploits the convex structure of the traffic assignment problem. At each iteration, the algorithm linearizes the objective function around the current flow pattern and identifies a descent direction by solving a simpler subproblem: computing shortest paths based on current link travel times and assigning all traffic to these paths in an all-or-nothing manner. This all-or-nothing assignment represents the flow pattern that would emerge if all travelers simultaneously chose their shortest paths given current conditions. The algorithm then performs a line search to determine the optimal step size $\alpha \in [0, 1]$ that defines a convex combination between the current flow pattern and the all-or-nothing auxiliary flows. This step size balances the benefit of moving toward routes that are currently shortest (based on the existing flow-dependent travel times) against the increased congestion that would result from such a shift. By iteratively updating flows in this manner, the algorithm converges to an equilibrium state where travel times are balanced across alternative routes for each origin-destination pair, satisfying the principle of user equilibrium. To accelerate convergence, our implementation uses the Conjugate Frank-Wolfe variant, which augments the all-or-nothing direction with information from previous iterations to form a conjugate descent direction, and terminates when the relative duality gap falls below a tolerance. The full algorithm is described in~\ref{appendix:frank_wolfe}.

The Frank-Wolfe algorithm must be executed for each time period $t \in \mathcal{T}$ to compute the traffic flows under the capacity reductions imposed by the renovation schedule determined at the upper level. These computed flows determine the link travel times $c_{ij}^t$ that feed into the $TTD$ objective in Equation~\ref{eq:ttd}. The computational expense of repeatedly solving traffic assignment problems across multiple time periods and candidate solutions motivates the progressive evaluation strategy introduced in Section~\ref{subsubsec:PLBE}, which reduces the number of traffic assignments required during optimization.

\section{Experiments}
\label{sec:experiments}
In this section, we evaluate PLBE through four experiments of increasing scale. Section~\ref{subsec:exp_setup} describes the common experimental setup. Section~\ref{subsec:exp1_new} validates the method on a small instance whose exact Pareto front is obtained by branch-and-bound; Section~\ref{subsec:exp2_new} measures headline computational efficiency on the full Sioux Falls instance; Section~\ref{subsec:exp3_new} tests robustness across systematically varied instances; and Section~\ref{subsec:exp4_new} examines application to the larger, more realistic Anaheim network. Section~\ref{subsec:interpret} then interprets the resulting trade-off for decision-makers.
\subsection{Experimental Setup}
\label{subsec:exp_setup}
\begin{figure*}[htbp]
    \centering
    \includegraphics[width=0.5\textwidth]{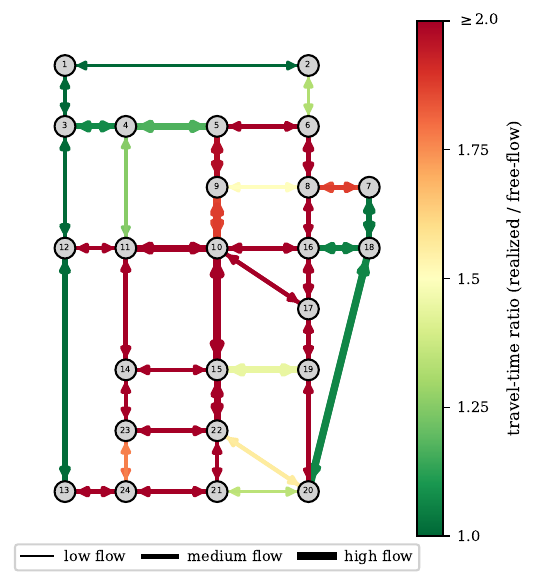}
    \caption{The Sioux Falls traffic network used in the experiments. Link colors indicate the ratio between realized and free-flow travel time, ranging from green (ratio = 1, no congestion) to red (ratio $\geq 2$ , severe congestion). Link thickness is proportional to traffic flow volume.}
    \label{fig:sf_network}
\end{figure*}

The experiments use four problem instances of increasing scale. The base instance, SF-76, is the well-established Sioux Falls network --- a standard test case in transportation optimization (see Table~\ref{table:lit_overview}) --- with one renovation project generated for each of its 76 directed links. Prior work on Sioux Falls typically considers far fewer projects (around ten; see Table~\ref{table:lit_overview}), so generating a project for every link yields a substantially larger instance than those typically studied in the literature. Figure~\ref{fig:sf_network} illustrates the topology. For validation against an exact solution (Experiment~1) we use a small instance, SF-12, with 12 projects, solvable exactly within a reasonable time budget. Robustness (Experiment~3) is assessed on a family of SF-76 variants that systematically modify congestion, constraints, and topology (Table~\ref{tab:case_mutations}). Experiment~4 tests application to the larger, more realistic Anaheim network, with 914 links and 80 data-driven congestion-corridor projects.
The problem parameters are configured to reflect realistic infrastructure planning scenarios. We divide the planning horizon into discrete time periods of one quarter (three months), spanning approximately 18 years for a total of 73 periods. Quarterly periods provide sufficient resolution to capture seasonal construction patterns while remaining computationally feasible. As discussed in Section~\ref{subsec:ULproblem}, each project is characterized by both a soft deadline reflecting cumulative failure probability and a hard deadline beyond which scheduling becomes infeasible. Risk parameters ($v_p$, $k_p$, and $w_p$ see Table~\ref{table:UL_notation}) are randomly generated for each project, producing heterogeneous risk profiles across the portfolio. Project durations are drawn from a Poisson distribution with mean $\mu = 5$ periods (15 months), reflecting typical infrastructure renovation time frames observed in practice.

Budget and capacity constraints are set to balance realism with computational feasibility. The construction budget accumulates at a fixed amount per period, totalling approximately 170\% of the sum of the project costs (to balance the restricting force of budgets with possible solution variety), and the maximum number of simultaneous projects is set at 9. We generate the problem instance with these characteristics, resulting in a total work sum of 356 construction-periods, meaning that on average 4.88 projects are active per time period. All algorithms use a fixed stopping criterion of 24 hours of wall-clock computation time, and each run is executed single-threaded on one core of the Dutch national supercomputer Snellius (``Rome'' thin nodes: two 64-core AMD EPYC 7H12 CPUs at 2.6\,GHz, 256\,GiB RAM); runs are dispatched as an array of independent single-core tasks. All traffic assignments use the Conjugate Frank--Wolfe solver with the relative-gap convergence criterion and renovation-crippling constants detailed in \ref{appendix:frank_wolfe}. A population size of 100 is used in all runs. We run each algorithm configuration 30 times with different random seeds due to the randomness inherent in NSGA-II, following standard practice in stochastic optimization for enabling statistical testing~\citep{derracPracticalTutorialUse2011}.

Across the experiments, we compare standard NSGA-II (\texttt{S|-|-})---the control, which evaluates every candidate schedule exactly with no surrogate pruning---against the PLBE variants --- Lazy Evaluation (LE) and Elimination Pruning (EP), each using the CostliestSubsetHeuristic (\texttt{H}) or XGBoost (\texttt{X}) surrogate trained with pinball loss at quantiles $q \in \{0.05, 0.1, 0.2, 0.5\}$ --- and a schedule-level surrogate-assisted baseline (\texttt{SS|X|0.5}). Experiment~1 additionally computes the exact Pareto front with a branch-and-bound solver (\texttt{BB}). Table~\ref{tab:algorithm_overview} summarizes the algorithm configurations and naming convention.

\begin{table}[htbp]
\centering
\caption{Overview of algorithm configurations. Algorithm naming convention: [Evaluator]|[Regressor]|[Quantile], where LE = Lazy Evaluation, EP = Elimination Pruning, H = Heuristic, X = XGBoost, S = Standard NSGA-II, SS = schedule-level surrogate-assisted baseline, and BB = the exact branch-and-bound solver (Experiment 1 only).}
\label{tab:algorithm_overview}
\small
\setlength{\tabcolsep}{4pt}
\begin{tabular}{lllc}
\toprule
\textbf{Algorithm} & \textbf{Evaluator} & \textbf{Regressor} & \textbf{Pinball Loss} \\
\midrule
LE|H|- & LazyEvaluation & Heuristic & -- \\
LE|X|0.05 & LazyEvaluation & XGBoost & 0.05 \\
LE|X|0.1 & LazyEvaluation & XGBoost & 0.1 \\
LE|X|0.2 & LazyEvaluation & XGBoost & 0.2\\
LE|X|0.5 & LazyEvaluation  & XGBoost & 0.5 \\
EP|H|- & EliminationPruning  & Heuristic & -- \\
EP|X|0.05 & EliminationPruning  & XGBoost & 0.05 \\
EP|X|0.1 & EliminationPruning & XGBoost  & 0.1\\
EP|X|0.2 & EliminationPruning & XGBoost&0.2\\
EP|X|0.5 & EliminationPruning  & XGBoost  & 0.5   \\
\midrule
SS|X|0.5 & ScheduleSurrogate & XGBoost & 0.5 \\
S|-|- & Standard     & None    & --   \\
BB & Branch-and-Bound & None & --   \\
\bottomrule
\end{tabular}
\end{table}

We assess each configuration on solution quality and computational cost. Hypervolume (HV) measures the dominated objective-space volume relative to a reference point. A reference point taken from the worst observed objective values saturates HV once fronts are dense, so in all experiments we place the reference point just beyond the realistic operating range of the instance at hand, keeping quality differences visible; for SF-76 this is $(R, TTD) = (2000,\, 2\times10^{9})$. We additionally report the modified inverted generational distance (IGD$^{+}$) against a pooled best-known front, the minimum normalized distance to the ideal point (Dist.), and the Pareto-front size (PF Size). We also report the two-set coverage $C(A,B)$: the fraction of the solutions in front $B$ dominated by at least one solution in front $A$, so that $C(A,B)=1$ means front $A$ dominates front $B$ entirely; coverage is computed per seed pair and reported as the median. Computational cost is captured by the number of unique traffic simulations (Unique Sims, the dominant expense) and the number of NSGA-II generations completed within the budget (Iter.). Differences across seeds are assessed with non-parametric tests (Friedman across configurations; Mann--Whitney against the control)~\citep{derracPracticalTutorialUse2011}.

\subsection{Experiment 1: Validation against the Exact Pareto Front (SF-12)}
\label{subsec:exp1_new}
SF-12 is deliberately compressed so that an exact front is reachable: it spans a 5-year horizon (20 quarterly periods), allows at most five simultaneous projects, and leaves the budget non-binding; project durations average five periods, giving roughly three concurrent projects per period. Even so, its nominal solution space is approximately $8\times10^{13}$ schedules, and computing the exact Pareto front by branch-and-bound (\texttt{BB}) required about 8 hours; SF-12 is thus the largest instance we could solve exactly within a reasonable budget. Against this exact front (95 non-dominated solutions), we evaluate all PLBE variants (Elimination Pruning and Lazy Evaluation, each with the XGBoost and heuristic surrogates at quantiles $q\in\{0.05,0.1,0.2,0.5\}$), standard NSGA-II (\texttt{S|-|-}), and the schedule-level SAEA baseline (\texttt{SS|X|0.5}), over 30 seeds. Table~\ref{tab:e1_results} reports the results.

All PLBE variants recover the exact front almost perfectly: the mean hypervolume gap is below $0.3\%$, statistically indistinguishable from exhaustive standard NSGA-II (Mann--Whitney on HV, all $p>0.05$). The schedule-level surrogate, by contrast, is significantly worse --- a $3.5\%$ hypervolume gap ($p\approx3\times10^{-11}$) and less than half the Pareto-front size. Predicting whole-schedule travel delay and pre-selecting on it biases the search away from the true front, whereas PLBE's conservative component-level lower-bound estimates preserve it; this motivates the progressive, component-wise pruning at the core of our method.

\begin{figure}[htbp]
    \centering
    \includegraphics[width=\columnwidth]{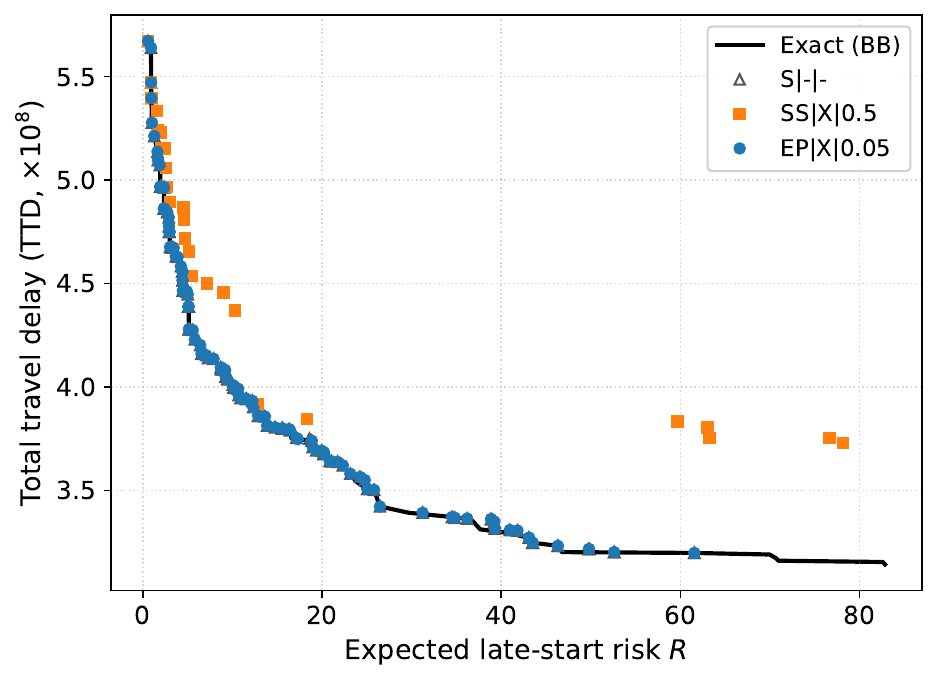}
    \caption{SF-12: recovered final Pareto fronts (representative median-hypervolume seed per method) for \texttt{EP|X|0.05}, the schedule-level SAEA baseline (\texttt{SS|X|0.5}), and standard NSGA-II (\texttt{S|-|-}), overlaid on the exact branch-and-bound front. PLBE and standard NSGA-II both recover the exact front closely, whereas the SAEA baseline departs from it (higher travel delay at matched risk, and scattered high-risk solutions).}
    \label{fig:e1_pareto}
\end{figure}

Because SF-12 is small, the simulation savings are modest: standard NSGA-II saturates the reachable scenario set at 1{,}586 unique simulations, while the PLBE variants converge to the same front using $\sim$1{,}020--1{,}190 simulations ($1.3$--$1.6\times$ fewer). This margin grows sharply with instance size (Experiments~2 and~4). Across the family, the surrogate quantile $q$ trades solution quality against efficiency: a lower quantile yields a more conservative lower bound that prunes less aggressively (the safer end of the rule), while a higher quantile prunes more and saves more simulations. We also measured directly how often the pruning rule discards a schedule that exact evaluation would have kept on the front, using a log-and-replay diagnostic that logs a random sample of pruned candidates together with the incumbent non-dominated set and re-evaluates them exactly. The false-pruning rate is small at both scales: \texttt{EP|X|0.05} wrongly prunes a truly non-dominated schedule in only $0.48\%$ of pruning decisions on SF-12 ($95\%$ Wilson CI $[0.46\%,\,0.50\%]$) and $0.19\%$ on SF-76 (CI $[0.16\%,\,0.23\%]$), confirming that the conservative lower bound rarely prunes incorrectly. Figure~\ref{fig:e1_pareto} overlays the recovered fronts on the exact front.

\begin{table*}[htbp]
\centering
\caption{Experiment 1 (SF-12): performance against the exact branch-and-bound Pareto front, mean (standard deviation) over 30 seeds. HV gap is relative to the exact front (lower is better). The exact solver \texttt{BB} is the reference and is omitted from the comparison rows.}
\label{tab:e1_results}
\begin{tabular}{l|cccccc}
\toprule
Algorithm & HV & HV gap (\%) & Min Dist & PF Size & Unique Sims & Iter. \\
\midrule
S|-|-     & 0.8395 (0.0018) & 0.291 (0.216) & 0.1627 (0.0013) & 93 (3) & 1586 (0)   & 24548 (5658) \\
SS|X|0.5  & 0.8125 (0.0029) & 3.494 (0.342) & 0.1907 (0.0029) & 38 (7) & 1201 (10)  & 47077 (5025) \\
\midrule
EP|H|-    & 0.8403 (0.0016) & 0.187 (0.190) & 0.1623 (0.0007) & 95 (3) & 1037 (11)  & 39158 (9895) \\
EP|X|0.05 & 0.8396 (0.0021) & 0.279 (0.248) & 0.1627 (0.0014) & 94 (3) & 1129 (43)  & 36627 (9673) \\
EP|X|0.1  & 0.8396 (0.0020) & 0.280 (0.236) & 0.1627 (0.0014) & 93 (3) & 1047 (22)  & 42994 (7013) \\
EP|X|0.2  & 0.8402 (0.0016) & 0.204 (0.193) & 0.1623 (0.0008) & 95 (3) & 1025 (8)   & 45666 (4185) \\
EP|X|0.5  & 0.8399 (0.0016) & 0.243 (0.185) & 0.1624 (0.0008) & 92 (3) & 1016 (7)   & 42851 (5956) \\
\midrule
LE|H|-    & 0.8397 (0.0018) & 0.257 (0.214) & 0.1626 (0.0012) & 94 (3) & 1058 (11)  & 28964 (7675) \\
LE|X|0.05 & 0.8401 (0.0018) & 0.214 (0.211) & 0.1624 (0.0011) & 94 (3) & 1191 (43)  & 28546 (7755) \\
LE|X|0.1  & 0.8396 (0.0016) & 0.281 (0.191) & 0.1625 (0.0010) & 93 (2) & 1087 (32)  & 34765 (9563) \\
LE|X|0.2  & 0.8398 (0.0019) & 0.254 (0.228) & 0.1626 (0.0012) & 94 (3) & 1046 (9)   & 38835 (8844) \\
LE|X|0.5  & 0.8396 (0.0018) & 0.274 (0.219) & 0.1627 (0.0013) & 91 (4) & 1036 (9)   & 42487 (7298) \\
\bottomrule
\end{tabular}
\end{table*}

\subsection{Experiment 2: Computational Efficiency at Scale (SF-76)}
\label{subsec:exp2_new}
This experiment measures headline efficiency at scale, comparing the PLBE family against standard NSGA-II (\texttt{S|-|-}) and the schedule-level SAEA baseline (\texttt{SS|X|0.5}) on the full Sioux Falls instance (SF-76: 76 projects), over 30 seeds within the 24-hour budget. Table~\ref{tab:e2_results} reports the results.

Within the fixed budget, \texttt{EP|X|0.05} completes roughly 87$\times$ more generations than standard NSGA-II, using 6.8$\times$ fewer unique traffic simulations and returning Pareto fronts 3.5$\times$ larger. Each of these differences is statistically significant across the 30 seeds.

\texttt{EP|X|0.05} attains significantly higher hypervolume than both standard NSGA-II and the SAEA baseline, which are themselves statistically indistinguishable; the advantage is more pronounced on IGD$^{+}$, where PLBE is again significantly better than both baselines. Two-set coverage corroborates this: PLBE dominates the entire standard-NSGA-II front and about 96\% of the SAEA front, with negligible coverage in the reverse direction (Table~\ref{tab:e2_results}). Based on these results, we adopt \texttt{EP|X|0.05} as the representative PLBE configuration for Experiments~3 and~4, prioritizing solution quality: it attains significantly higher hypervolume and lower IGD$^{+}$ than higher quantiles, which in turn trade a small quality margin for roughly 15--16\% additional simulation reduction.

\begin{figure*}[htbp]
    \centering
    \includegraphics[width=\textwidth]{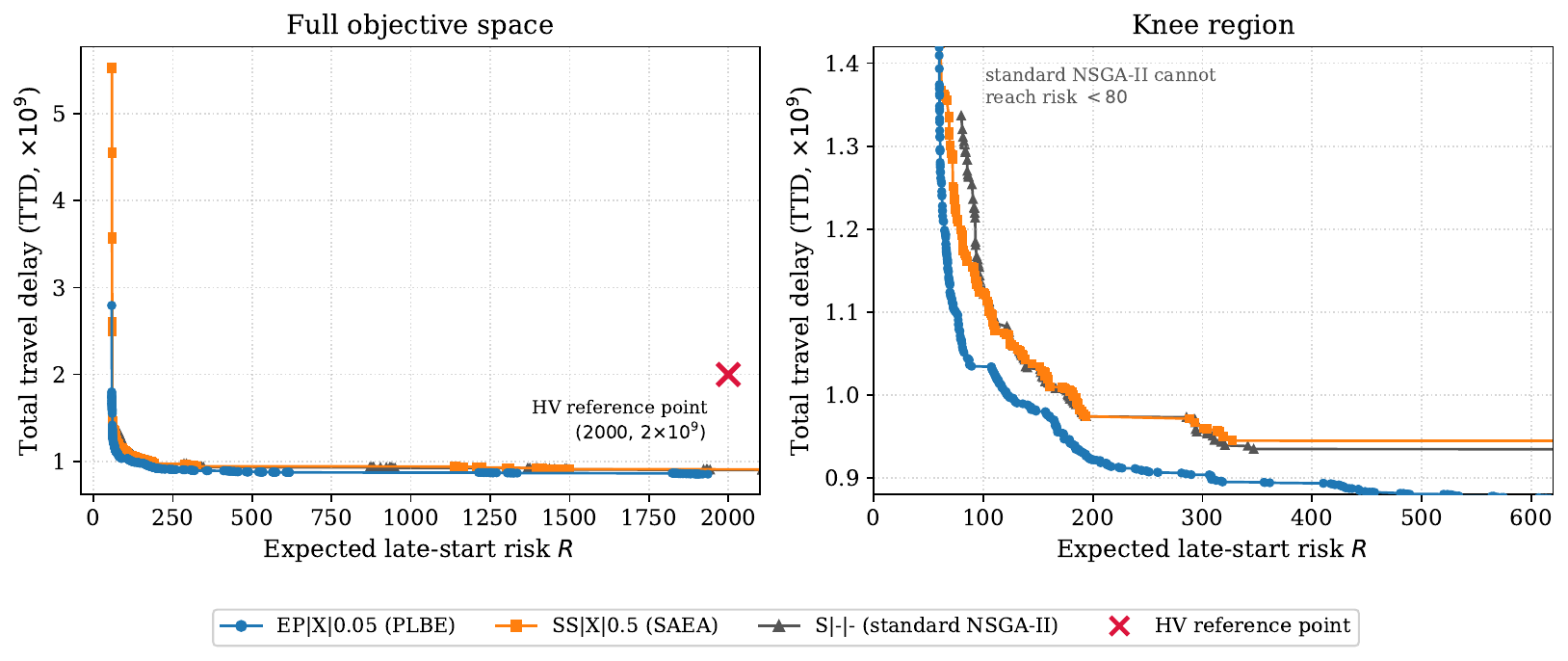}
    \caption{SF-76: representative final Pareto fronts (the median-hypervolume seed per method; for \texttt{EP|X|0.05}, the run also used in Section~\ref{subsec:interpret}) for \texttt{EP|X|0.05}, the schedule-level SAEA baseline (\texttt{SS|X|0.5}), and standard NSGA-II (\texttt{S|-|-}), in (Risk, $TTD$) space. Left: full objective space with the hypervolume reference point marked. Right: knee region. PLBE produces a denser, better-converged front that reaches lower-risk solutions than either baseline.}
    \label{fig:e2_pareto}
\end{figure*}

\begin{figure*}[htbp]
    \centering
    \includegraphics[width=\textwidth]{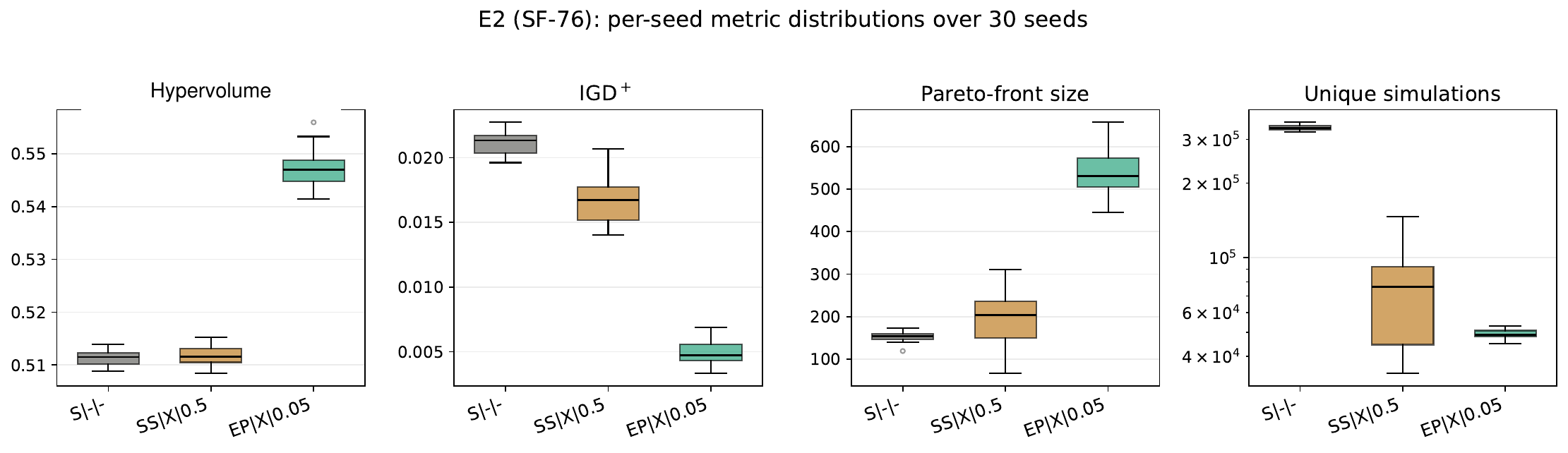}
    \caption{SF-76: per-seed distributions (30 seeds) of the discriminating quality and cost metrics --- hypervolume, IGD$^{+}$, Pareto-front size, and unique simulations (log scale) --- for \texttt{EP|X|0.05}, the schedule-level SAEA baseline (\texttt{SS|X|0.5}), and standard NSGA-II (\texttt{S|-|-}).}
    \label{fig:e2_boxplots}
\end{figure*}

Notably, the schedule-level SAEA baseline is not competitive with PLBE: it is statistically indistinguishable from standard NSGA-II and is almost entirely dominated by PLBE. It does reduce simulations relative to the control (about 4.5$\times$), but it matches neither PLBE's efficiency nor its solution quality, reinforcing that the benefit comes from component-level progressive pruning rather than from surrogate assistance per se. Figure~\ref{fig:e2_pareto} shows representative fronts and Figure~\ref{fig:e2_boxplots} the per-seed distributions of the discriminating metrics.

\begin{table*}[htbp]
\centering
\caption{Experiment 2 (SF-76): mean (standard deviation) over 30 seeds. Hypervolume is computed against the reference point $(R,TTD)=(2000,\,2\times10^{9})$; IGD$^{+}$ is measured against the pooled best-known front. Higher HV / PF Size and lower IGD$^{+}$ / Min Dist / Unique Sims are better.}
\label{tab:e2_results}
\begin{tabular}{l|cccccc}
\toprule
Algorithm & HV & IGD$^{+}$ & Min Dist & PF Size & Unique Sims & Iter. \\
\midrule
S|-|-     & 0.5113 (0.0015) & 0.0211 (0.0009) & 0.4966 (0.0009) & 153 (11)  & 333540 (9059)  & 153 (4)      \\
SS|X|0.5  & 0.5118 (0.0018) & 0.0167 (0.0018) & 0.4972 (0.0012) & 189 (58)  & 73344 (29958)  & 9373 (2549)  \\
\midrule
EP|X|0.05 & 0.5471 (0.0033) & 0.0048 (0.0009) & 0.4685 (0.0047) & 538 (49)  & 49204 (1960)   & 13312 (659)  \\
EP|X|0.2  & 0.5446 (0.0026) & 0.0055 (0.0011) & 0.4709 (0.0047) & 531 (63)  & 41383 (2473)   & 13290 (1202) \\
EP|X|0.5  & 0.5358 (0.0029) & 0.0079 (0.0010) & 0.4810 (0.0047) & 493 (59)  & 35940 (1443)   & 10721 (2369) \\
LE|X|0.05 & 0.5464 (0.0034) & 0.0053 (0.0011) & 0.4692 (0.0046) & 419 (45)  & 61863 (5295)   & 12696 (670)  \\
LE|X|0.2  & 0.5437 (0.0047) & 0.0060 (0.0011) & 0.4724 (0.0045) & 418 (48)  & 45022 (2056)   & 12692 (1881) \\
LE|X|0.5  & 0.5373 (0.0052) & 0.0081 (0.0012) & 0.4785 (0.0057) & 371 (41)  & 38301 (2036)   & 9426 (1587)  \\
\bottomrule
\end{tabular}
\end{table*}

\subsection{Experiment 3: Robustness across Problem Variants}
\label{subsec:exp3_new}
This experiment tests whether the efficiency and quality gains of PLBE persist when the problem is systematically varied along three axes: network congestion (road-capacity scaling), construction-resource tightness (simultaneous-project and budget limits), and network topology (links added or removed). For each variant we compare the representative PLBE configuration (\texttt{EP|X|0.05}) against standard NSGA-II (\texttt{S|-|-}) and the schedule-level SAEA baseline (\texttt{SS|X|0.5}), using the common 30-seed, 24-hour protocol of Section~\ref{subsec:exp_setup}. Table~\ref{tab:case_mutations} summarizes the seven variants, and the two topology modifications are illustrated in Appendix~\ref{appendix:network_variants} (Figure~\ref{fig:network_variants}).

The three road-capacity variants rescale every link capacity, sharpening or softening the congestion that closures induce: at $0.9\times$ the network is more congestion-sensitive, at $1.1\times$ less so, and at $100\times$ congestion is effectively removed (a near free-flow regime). The two construction-resource variants tighten or relax the scheduling constraints: the $0.7\times$ variant permits at most six simultaneous projects (versus nine) with a proportionally smaller per-period budget, making feasible schedules harder to find, while the $100\times$ variant effectively removes both limits. The two topology variants remove or add links to lower or raise connectivity, with link capacities adjusted to keep the baseline assignment well defined. Together these span the congestion regimes, constraint pressures, and connectivity patterns a planner is likely to encounter.

\begin{table*}[htbp]
\centering
\caption{Overview of the seven SF-76 problem variants used in Experiment 3.}
\label{tab:case_mutations}
\begin{tabularx}{\textwidth}{>{\raggedright\arraybackslash}p{3.6cm} X}
\toprule
\textbf{Variant} & \textbf{Description} \\
\midrule
Reduced road capacity ($0.9\times$) & All link capacities scaled to 90\% of their base value, increasing congestion and the travel-delay impact of road closures. \\
\addlinespace[0.4em]
Increased road capacity ($1.1\times$) & All link capacities scaled to 110\% of their base value, reducing congestion effects. \\
\addlinespace[0.4em]
Maximal road capacity ($100\times$) & Link capacities scaled by a factor of 100, effectively eliminating congestion (near free-flow conditions). \\
\addlinespace[0.4em]
Tight construction capacity ($0.7\times$) & At most six simultaneous projects (versus nine) and a per-period budget scaled to 0.7$\times$, tightening the scheduling constraints. \\
\addlinespace[0.4em]
Relaxed construction capacity ($100\times$) & Simultaneous-project and per-period budget limits scaled by 100$\times$, effectively removing the construction-resource constraints. \\
\addlinespace[0.4em]
Less connected network & Links removed to reduce connectivity (Figure~\ref{fig:sioux_falls_less_connected}); link capacities raised to compensate. \\
\addlinespace[0.4em]
More connected network & Links added to increase connectivity (Figure~\ref{fig:sioux_falls_more_connected}); link capacities lowered to compensate. \\
\bottomrule
\end{tabularx}
\end{table*}

Table~\ref{tab:e3_results} reports the outcome; the underlying per-seed distributions are shown in Figure~\ref{fig:e3_boxplots} (Appendix~\ref{appendix:network_variants}). In every variant, \texttt{EP|X|0.05} dominates the entire standard-NSGA-II front ($C=1.00$, none in reverse) and attains significantly higher hypervolume and lower IGD$^{+}$ (Friedman $p<10^{-6}$). It uses $3$--$17\times$ fewer unique simulations than standard NSGA-II and returns $1.5$--$3.9\times$ larger fronts; the simulation saving is largest ($17\times$) under the tightest construction capacity. The SAEA baseline is dominated $83$--$100\%$ throughout. Four $\Delta$HV differences in Table~\ref{tab:e3_results} are not statistically significant ($\dagger$); one of them is negative---under tight construction capacity---and this single nominal edge of the baseline occurs exactly where its aggressive pre-selection uses the fewest simulations, as the table's best-baseline ratio columns show. Since hypervolume saturates for the low-congestion variants, the cross-variant comparison rests on the reference-independent metrics.

\begin{table*}[htbp]
\centering
\caption{Experiment 3 (SF-76 variants): PLBE (\texttt{EP|X|0.05}) against the better of the two baselines (standard NSGA-II \texttt{S|-|-} or SAEA \texttt{SS|X|0.5}) on each metric, over 30 seeds per variant. $\Delta$HV is PLBE's mean hypervolume minus that of the best baseline ($\dagger$: difference not statistically significant); $C(\mathrm{EP},\cdot)$ is the median two-set coverage of each baseline's front by PLBE (reverse coverages are at most $0.01$). The IGD$^{+}$, Pareto-front-size, and unique-simulation columns report PLBE relative to whichever baseline performs best on that metric, so values $>1$ favor PLBE.}
\label{tab:e3_results}
\begin{tabular}{l|cccccc}
\toprule
Variant & $\Delta$HV & $C$(EP,S) & $C$(EP,SS) & IGD$^{+}$ ($\times$) & PF size ($\times$) & Sims ($\times$) \\
\midrule
Road capacity $0.9\times$         & $+0.0003^{\dagger}$ & 1.00 & 1.00 & 2.7 & 3.0 & 2.2 \\
Road capacity $1.1\times$         & $+0.0003^{\dagger}$ & 1.00 & 0.94 & 2.3 & 3.2 & 2.6 \\
Road capacity $100\times$         & $+0.0010$           & 1.00 & 0.94 & 2.2 & 2.4 & 1.2 \\
Construction capacity $0.7\times$ & $-0.0019^{\dagger}$ & 1.00 & 0.83 & 0.9 & 1.5 & 0.7 \\
Construction capacity $100\times$ & $+0.0015$           & 1.00 & 1.00 & 2.3 & 3.2 & 0.7 \\
Less connected                    & $+0.0002$           & 1.00 & 0.99 & 2.4 & 2.4 & 4.4 \\
More connected                    & $+0.0003^{\dagger}$ & 1.00 & 1.00 & 3.3 & 3.6 & 2.4 \\
\bottomrule
\end{tabular}
\end{table*}

\subsection{Experiment 4: Application to a Larger Network and Size Scaling (Anaheim-80)}
\label{subsec:exp4_new}
\begin{figure}[htbp]
    \centering
    \includegraphics[width=\columnwidth]{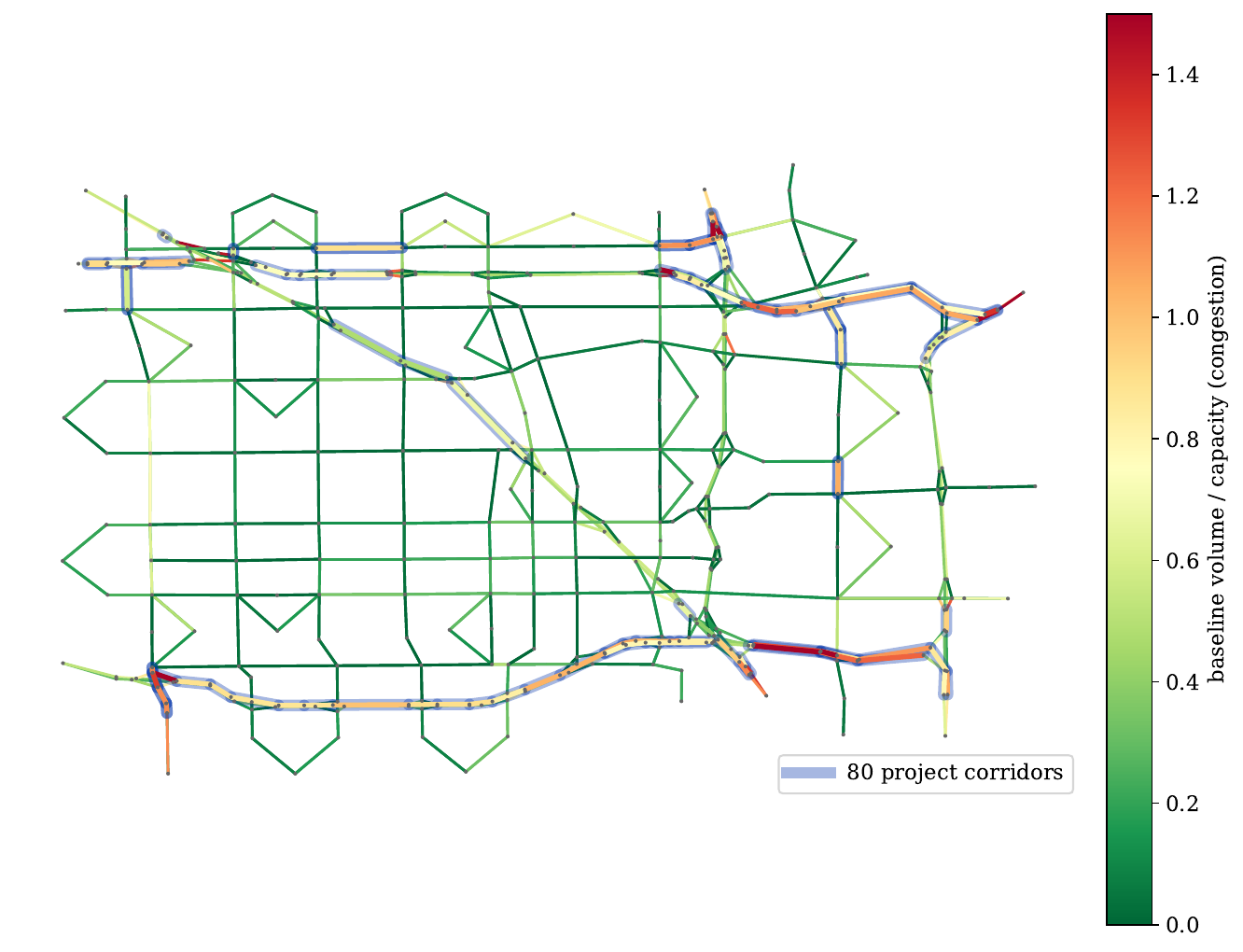}
    \caption{The Anaheim network used in Experiment 4. Links are colored by baseline volume-to-capacity ratio (green = uncongested through red = congested); the 80 renovation-project corridors, selected as the highest-impact corridors, are highlighted.}
    \label{fig:anaheim}
\end{figure}

The final experiment tests whether PLBE extends to a substantially larger and more realistic network. We use the Anaheim network (914 links, 416 nodes, 38 zones; Figure~\ref{fig:anaheim}), with 80 renovation projects (composed of 112 directed links in total) derived data-drivenly as the highest-impact road corridors, ranked by mean volume-to-capacity ratio times total volume on a baseline user-equilibrium assignment, so that the selected corridors are both congested and heavily used. As in Experiments~2 and~3 we compare \texttt{EP|X|0.05} against standard NSGA-II (\texttt{S|-|-}) and the schedule-level SAEA baseline (\texttt{SS|X|0.5}) under the same 30-seed, 24-hour protocol used for SF-76. Because each traffic assignment on Anaheim is considerably more expensive than on Sioux Falls, all methods complete fewer generations here than on SF-76.

\begin{figure}[htbp]
    \centering
    \includegraphics[width=\columnwidth]{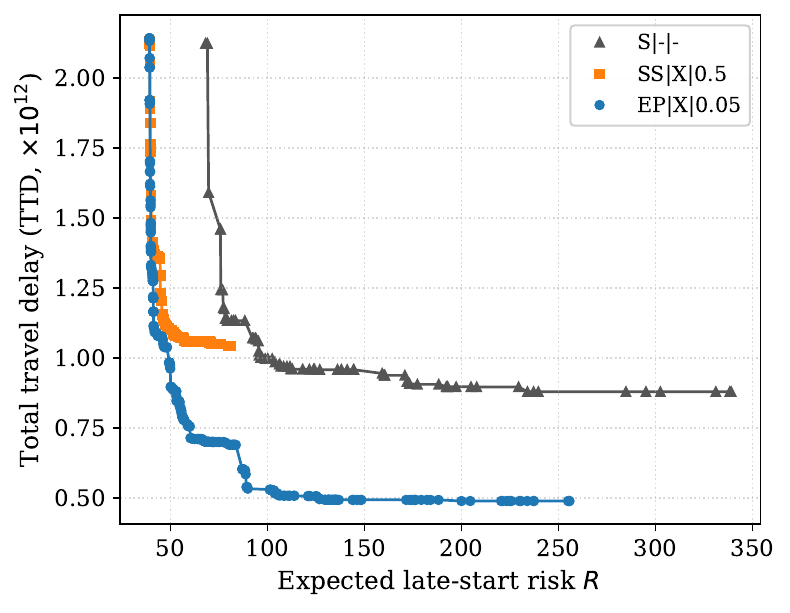}
    \caption{Anaheim-80: representative final Pareto fronts (the median-hypervolume seed per method) for \texttt{EP|X|0.05}, the schedule-level SAEA baseline (\texttt{SS|X|0.5}), and standard NSGA-II (\texttt{S|-|-}), in (Risk, $TTD$) space. PLBE reaches substantially lower travel delay across the whole risk range and dominates both baselines.}
    \label{fig:e4_pareto}
\end{figure}

Table~\ref{tab:e4_results} reports the results. The result is unambiguous: \texttt{EP|X|0.05} attains a mean hypervolume of $0.746$ against standard NSGA-II's $0.569$---a $31\%$ relative improvement---with far lower IGD$^{+}$ and roughly seven times as many Pareto solutions (Table~\ref{tab:e4_results}). Two-set coverage is complete --- PLBE dominates the entire standard-NSGA-II front with no coverage in the reverse direction --- and all differences are statistically significant (Friedman $p<10^{-10}$). Within the budget PLBE completes 43 times as many generations as standard NSGA-II while using $2.0\times$ fewer unique traffic simulations. As at SF-76, the schedule-level SAEA baseline is not competitive: despite pruning the most aggressively (the fewest simulations of the three), its hypervolume ($0.505$) falls below even standard NSGA-II's, reaffirming that the gain stems from component-level progressive pruning rather than from surrogate assistance per se. Figure~\ref{fig:e4_pareto} shows representative final Pareto fronts.

\begin{table*}[htbp]
\centering
\caption{Experiment 4 (Anaheim-80): mean (standard deviation) over 30 seeds. Hypervolume is computed against an Anaheim-specific reference point just beyond the pooled nadir of the observed fronts, $(R,TTD)\approx(1.2\times10^{3},\,2.25\times10^{12})$; IGD$^{+}$ is measured against the pooled best-known front. Higher HV / PF Size and lower IGD$^{+}$ / Min Dist / Unique Sims are better.}
\label{tab:e4_results}
\begin{tabular}{l|cccccc}
\toprule
Algorithm & HV & IGD$^{+}$ & Min Dist & PF Size & Unique Sims & Iter. \\
\midrule
S|-|-     & 0.5686 (0.0261) & 0.2867 (0.0401) & 0.4338 (0.0333) & 60 (13)  & 243457 (6994)  & 194 (31)     \\
SS|X|0.5  & 0.5050 (0.0714) & 0.2650 (0.0738) & 0.4807 (0.0727) & 153 (51) & 55312 (23222)  & 10785 (1201) \\
\midrule
EP|X|0.05 & 0.7462 (0.0322) & 0.0545 (0.0285) & 0.2515 (0.0309) & 425 (65) & 121949 (18783) & 8301 (987)   \\
\bottomrule
\end{tabular}
\end{table*}

Read across the three instances, PLBE's advantage over standard NSGA-II widens with problem size, and does so on reference-independent measures. On the small SF-12 instance, where standard NSGA-II nearly exhausts the reachable search space, the two are statistically indistinguishable (Section~\ref{subsec:exp1_new}). At SF-76 and on Anaheim-80, PLBE instead dominates the entire standard-NSGA-II front (two-set coverage $1.00$ in both cases), with an IGD$^{+}$ advantage that widens from roughly fourfold at SF-76 to more than fivefold on Anaheim, and Pareto fronts that grow from parity on SF-12 to $3.5\times$ larger at SF-76 and sevenfold on Anaheim. The unique-simulation reduction is substantial at every scale ($1.4\times$ on SF-12, $6.8\times$ on SF-76, $2.0\times$ on Anaheim-80) but not monotone in network size: on Anaheim each assignment is so costly that standard NSGA-II itself completes far fewer simulations, compressing the ratio. The decisive and monotone effect is on solution quality --- precisely in the large, congested regime where the exhaustive baseline is least able to cope, and which matters most for realistically sized planning problems.

\subsection{Objective Trade-off Analysis}
\label{subsec:interpret}
\begin{figure*}[htbp]
    \centering
    \includegraphics[width=\textwidth]{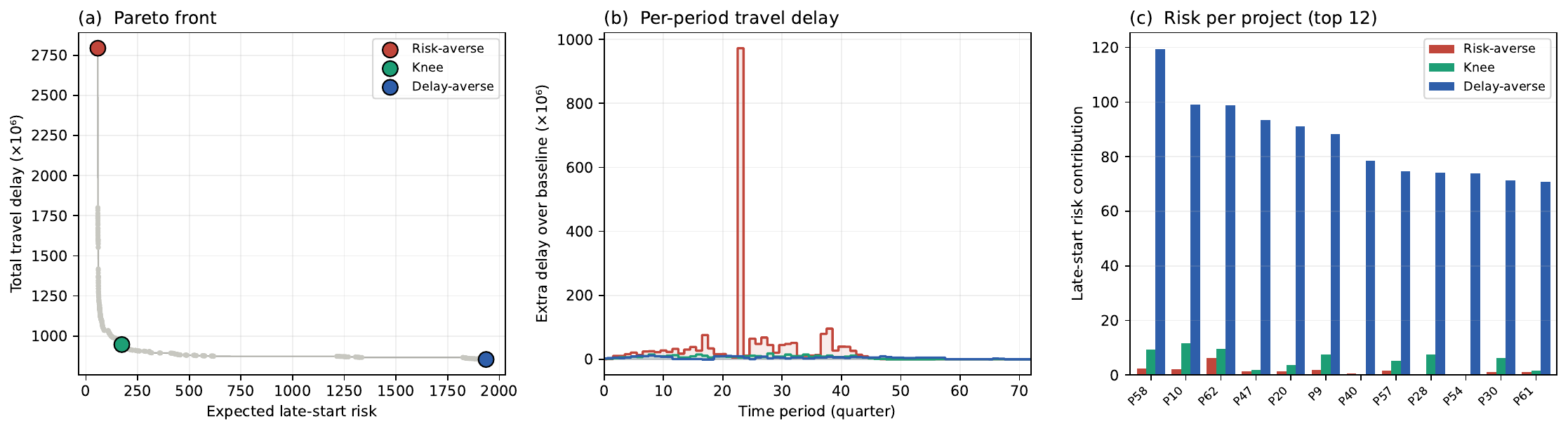}
    \caption{SF-76 representative run (\texttt{EP|X|0.05}): the risk--delay trade-off and where its cost arises. (a) Final Pareto front with three selected schedules. (b) Per-period travel delay over baseline for each. (c) Per-project late-start risk for the 12 highest-contributing projects.}
    \label{fig:interpret_tradeoff}
\end{figure*}

To illustrate what PLBE produces for a decision-maker, Figure~\ref{fig:interpret_tradeoff}(a) shows a representative final Pareto front (the \texttt{EP|X|0.05} run of Figure~\ref{fig:e2_pareto}) with three schedules: the risk-averse extreme (minimum risk), the delay-averse extreme (minimum travel delay), and the knee (closest to the ideal point). The front is sharply knee-shaped: moving from the delay-averse extreme to the knee reduces expected late-start risk roughly elevenfold for only about 11\% more travel delay, whereas pushing on to the risk-averse extreme then triples travel delay for a further threefold risk reduction. The knee thus delivers most of the achievable risk reduction at near-minimal travel delay. This knee shape arises because most of the risk reduction comes from starting a few high-cost projects early, which adds little travel delay, whereas the last increments of risk reduction require closing costly corridors concurrently.

Panels (b) and (c) show where these costs arise. The per-period travel delay (b) reveals that the risk-averse schedule's large delay is concentrated in a single period (around period~23), where a particularly costly combination of high-impact corridors closes at once; the knee and delay-averse schedules keep every period low by avoiding that combination. Travel delay therefore depends more on which corridors close at the same time than on how many close at once --- reflecting the non-additive coupling of closures through the network (Section~\ref{sec:problem}) that PLBE's per-scenario evaluation is built to exploit. The per-project risk (c) shows a similar pattern: the delay-averse schedule loads its risk onto a handful of high-cost projects deferred to the brink of their deadlines, whereas the knee starts exactly those projects early (near-zero risk) while sequencing the remainder to avoid the congestion spike.

\begin{figure}[htbp]
    \centering
    \includegraphics[width=\columnwidth]{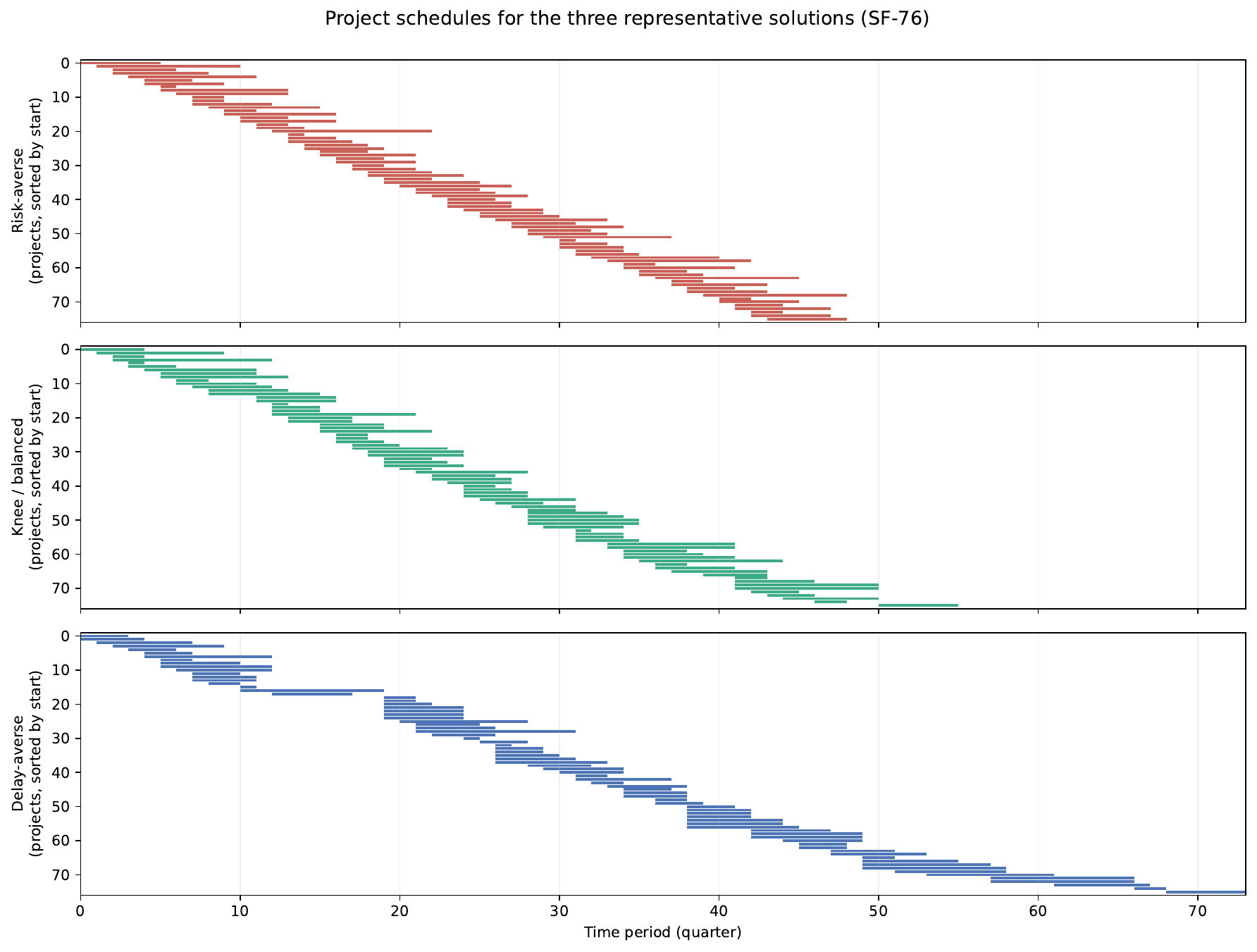}
    \caption{Project schedules (Gantt, sorted by start) for the three representative solutions: risk-averse compresses the portfolio into the early horizon, delay-averse spreads it across all 73 periods, the knee balances the two.}
    \label{fig:interpret_gantts}
\end{figure}

Figure~\ref{fig:interpret_gantts} shows the schedules themselves as Gantt charts. All three are bound by the nine-team construction-capacity limit, so they differ not in peak concurrency but in how they spread the same workload over the horizon: the risk-averse schedule compresses all projects into roughly the first 50 periods, the delay-averse schedule spreads them across the full 73-period horizon to thin out concurrent closures, and the knee packs the work almost as tightly as the risk-averse one while arranging which projects overlap.

\begin{figure}[htbp]
    \centering
    \includegraphics[width=\columnwidth]{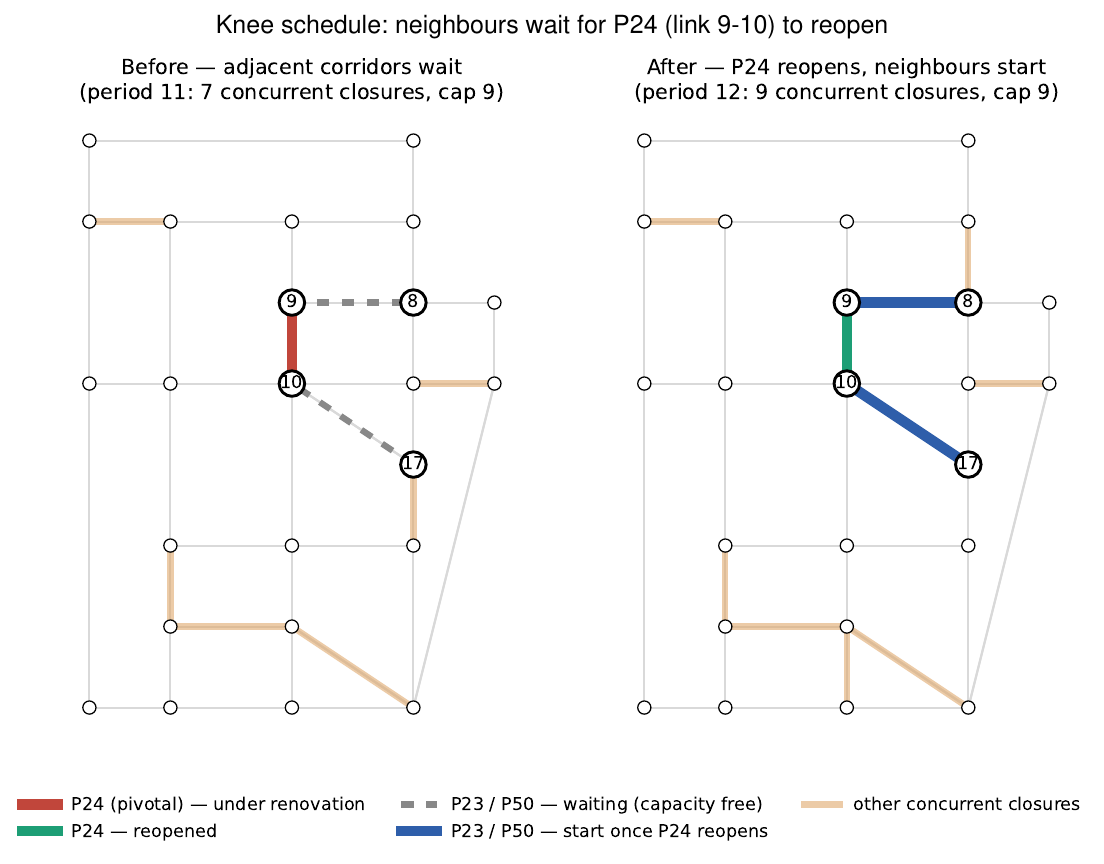}
    \caption{Learned sequencing in the knee schedule. Project P24 (link 9--10) is under renovation through period~11; its adjacent corridors P23 and P50---which share an endpoint (node 9) with P24---stay idle despite spare construction capacity and begin only once P24 reopens in period~12, avoiding three simultaneous closures around it.}
    \label{fig:interpret_sequencing}
\end{figure}

Closer inspection of the knee schedule shows that this arrangement includes learned sequencing around network bottlenecks. Figure~\ref{fig:interpret_sequencing} gives one illustrative example: project P24, which closes link 9--10 (incident to the most-connected node in the network), is under renovation through period~11; its two neighbouring corridors, P23 and P50, which share an endpoint with P24, remain idle even though two construction teams are free, and begin only once P24 reopens in period~12. Rather than close P24 together with both neighbours at once, which would sharply cut capacity around link 9--10, the optimizer defers the neighbours until after P24 finishes, trading a little late-start risk for a large reduction in concurrent congestion, recovering an implicit ``wait until the adjacent corridor reopens'' precedence that a planner would otherwise impose by hand. For a planner, then, the knee schedule is a natural operating point: it secures nearly all of the achievable risk reduction at close to the minimum travel delay, while the surrounding Pareto front quantifies the cost of moving away from it in either direction. What ultimately matters is not how many projects run at once---the construction-capacity limit fixes that---but which projects overlap, and in what order. Making that combinatorial structure affordable to search is precisely what PLBE enables.

\section{Conclusions}
\label{sec:conclusion}
This paper addressed the computational challenges of solving large-scale Road Network Maintenance Scheduling Problems with Uncertain Deadlines (RNMSP-UD) by introducing Progressive Lower Bound Evaluation (PLBE) methods integrated with NSGA-II. The proposed approach successfully tackles the exponential growth in computational requirements that arises when scheduling infrastructure renovations across urban networks while accounting for traffic impacts and uncertain infrastructure lifespans.

Our experimental evaluation on the Sioux Falls network with 76 renovation projects—substantially larger than the typical 5–35 projects found in existing literature---\allowbreak demonstrates that PLBE achieves an 87$\times$ improvement in iteration throughput compared to standard NSGA-II. Within a fixed 24-hour computational budget, XGBoost-based PLBE variants completed roughly 13{,}300 iterations compared to just 153 iterations for the baseline algorithm. This acceleration directly translates into superior solution quality: PLBE achieved statistically significant improvements in Hypervolume and Distance-to-origin metrics, with Pareto fronts containing up to 3.5 times as many non-dominated solutions. Regarding the surrogate, we adopted XGBoost trained on Quantile loss, which achieved the lowest prediction error among the models evaluated in our companion study \citep{boschMachineLearningPredictions2025}; on the small exactly-solvable instance even the training-free CostliestSubsetHeuristic recovers the exact front (Table~\ref{tab:e1_results}), and the specific quantile parameter value did not have a large effect on performance. When using XGBoost, Lazy Evaluation and Elimination Pruning achieved comparable performance, though Elimination Pruning showed greater robustness when paired with less accurate surrogates.

The sensitivity analysis across seven problem variants confirmed that these advantages are not artifacts of the base case configuration. Across diverse network topologies, constraint structures, and congestion intensities, PLBE consistently achieved 67--94\% reductions in required traffic simulations (a $3$--$17\times$ reduction) and produced Pareto fronts $1.5$--$3.9$ times larger than standard NSGA-II. Solution quality metrics showed consistent improvements with non-overlapping confidence intervals in most variants. The robustness of these gains across problem characteristics supports the method's applicability to the diverse conditions encountered in real urban infrastructure scheduling contexts.

From a theoretical perspective, this work contributes to the multi-objective optimization literature by demonstrating how machine learning can effectively approximate expensive black-box evaluations in population-based algorithms: the small quality cost of rare pruning errors is more than compensated by the efficiency gained, producing a significant net improvement in solution quality over the whole front. The progressive evaluation framework introduced here is generalizable to other domains where solution evaluation involves computationally expensive simulations.

For practitioners, our approach enables strategic level infrastructure planning at scales previously considered computationally intractable. The ability to evaluate schedules for up to 80 projects on a realistic 914-link urban network---where the method's advantage over standard NSGA-II is largest---while accounting for network-wide traffic impacts and uncertain infrastructure lifespans provides decision-makers with a more comprehensive methodology for balancing infrastructure reliability against traffic disruption costs. The multi-objective framework generates diverse solution sets that illuminate trade-offs between minimizing failure risk and reducing traffic congestion, supporting informed stakeholder planning. Concretely, a planning authority can run PLBE on its renovation portfolio within a day and present the resulting Pareto front to stakeholders: the knee schedule identified in Section~\ref{subsec:interpret} secures nearly all achievable risk reduction at close to minimal travel delay, while the surrounding front quantifies the cost of prioritizing either objective more strongly---turning schedule selection from a manual exercise into a transparent, evidence-based decision.

Future research directions include extending the approach to incorporate additional real-world complexities such as time-varying traffic demands and other traffic modalities. A further direction is to broaden the objective space to reflect the wider set of stakeholders affected by road renovation---incorporating, for example, pedestrian safety, local accessibility, and construction-related emissions as additional simulation-derived objectives. Because the progressive-evaluation framework operates on any objective that decomposes into per-scenario simulation outcomes, such objectives could in principle be added without altering the underlying pruning mechanism. Decision-making could be extended to include life-extending measures such as vehicle weight limits or temporary scaffolding; since these instruments often fall to separate departments or the contractor, such an extension would require institutional cooperation, though in other contexts it has precedent---Amsterdam, for example, imposes both on aging quay walls to defer reconstruction~\citep{nosKademuurCentrumAmsterdam2020}. The surrogate modeling approach could be enhanced through alternative machine learning methods. Finally, the static scheduling framework presented here could be extended to dynamic settings that incorporate new information about infrastructure degradation as it becomes available, potentially formulated as a Markov Decision Process or a two-stage stochastic optimization problem.

%% Required Sections for CACAIE %%

\section*{Acknowledgements}
This work was supported by the Netherlands Organisation for Scientific Research (NWO) through the Logiquay project (grant number NWA.1431.20.005).

\section*{Funding}
This research was funded by the Netherlands Organisation for Scientific Research (NWO) under grant number NWA.1431.20.005 (Logiquay project).

\section*{Declaration of Competing Interests}
The authors declare that they have no known competing financial interests or personal relationships that could have appeared to influence the work reported in this paper.

\section*{CRediT Author Contribution Statement}
\textbf{Robbert Bosch:} Conceptualization, Methodology, Software, Validation, Formal analysis, Investigation, Data curation, Writing -- original draft, Writing -- review \& editing, Visualization.
\textbf{Patricia Rogetzer:} Conceptualization, Writing -- review \& editing, Supervision, Funding acquisition.
\textbf{Wouter van Heeswijk:} Conceptualization, Writing -- review \& editing, Supervision, Project administration, Funding acquisition.
\textbf{Martijn Mes:} Conceptualization, Resources, Writing -- review \& editing, Supervision, Funding acquisition.

\section*{Data Availability}
The source code implementing the proposed methods, together with the generated problem instances and the analysis scripts used to produce the reported results, is openly available in a versioned, citable Zenodo archive (version~1.1.0; \url{https://doi.org/10.5281/zenodo.20640178}; development repository: \url{https://github.com/rrbosch/road-maintenance-scheduling}). The Sioux Falls and Anaheim network data used in this study are publicly available from the Transportation Networks repository (\url{https://github.com/bstabler/TransportationNetworks}).

\section*{Declaration of Generative AI and AI-assisted Technologies}
During the preparation of this work the author(s) used Claude (Anthropic) to improve the language and readability of author-provided content. After using this tool, the author(s) reviewed and edited the content as needed and take(s) full responsibility for the content of the publication.

\appendix

\newpage
\section{Upper Level Problem Formulation}
\label{appendix:UL_formulation}

This appendix provides the complete notation and constraints for the upper level scheduling problem formulated in Section~\ref{subsec:ULproblem}.

\begin{table*}[h!]
\centering
\caption{Main notation for the upper level scheduling problem}
\label{table:UL_notation}
\begin{tabularx}{\textwidth}{p{2cm}X}
\toprule
\multicolumn{2}{l}{\textbf{Objectives}} \\
$R$ & Expected cost of infrastructure failure \\
$TTD$ & Total increase in network travel time due to ongoing renovations, relative to a base scenario \\
$f_{ij}^t$, $c_{ij}^t$ & Equilibrium traffic flow and travel time on arc $(i,j) \in \mathcal{A}$ during period $t$ (determined by the lower level problem) \\
$tt_{\text{base}}$ & Flow-weighted total system travel time in the undisturbed network \\
\\
\multicolumn{2}{l}{\textbf{Sets and Parameters}} \\
$T$ & Set of discrete time periods $t \in \mathcal{T}$ \\
$\mathcal{P}$ & Set of renovation projects $p \in \mathcal{P}$ \\
$b_t$ & Budget available in time period $t$ \\
$m$ & Maximum number of projects that may be ongoing simultaneously \\
\\
\multicolumn{2}{l}{\textbf{Project-Specific Parameters (for $p \in \mathcal{P}$)}} \\
$l_p$ & Hard due date (latest permissible completion time) \\
$d_p$ & Duration of the project \\
$c_p$ & Cost of the project \\
$w_p$ & Cost associated with failure of the project \\
$v_p$ & Probability parameter for failure in binomial decay model \\
$k_p$ & Deterioration events tolerated before failure (failure once this is exceeded), in the binomial decay model \\
\\
\multicolumn{2}{l}{\textbf{Decision Variables}} \\
$X_{tp} \in \{0,1\}$ & 1 if project $p$ starts in period $t$, 0 otherwise \\
$G_{tp} \in \{0,1\}$ & 1 if project $p$ is ongoing in period $t$, 0 otherwise \\
\bottomrule
\end{tabularx}
\end{table*}

Each project must start exactly once, as enforced by constraint~\ref{eq:sum_x}: 
\begin{equation}\label{eq:sum_x} 
\sum_{t \in \mathcal{T}} X_{tp} = 1 \qquad \forall p \in \mathcal{P}
\end{equation}

Constraint~\ref{eq:x_zero} ensures that no project can start too late to finish before its hard deadline: 
\begin{equation}\label{eq:x_zero} 
X_{tp} = 0 \qquad \forall p \in \mathcal{P}, \forall t \geq l_p - d_p 
\end{equation}

Constraint~\ref{eq:G_start} specifies that a project is ongoing in a time period if it started in that period: 
\begin{equation}\label{eq:G_start} 
G_{tp} \geq X_{tp} \qquad \forall p \in \mathcal{P}, \forall t \in \mathcal{T} 
\end{equation}

Constraint~\ref{eq:G_continuation} ensures that once a project has started, it remains ongoing until completion: 
\begin{equation}\label{eq:G_continuation} 
G_{tp} \leq G_{t-1,p} + X_{tp} \qquad \forall p \in \mathcal{P}, \forall t \in \mathcal{T} \setminus \{0\} 
\end{equation}

Constraint~\ref{eq:G_total} enforces that each project is ongoing for exactly its specified duration: 
\begin{equation}\label{eq:G_total} 
\sum_{t \in \mathcal{T}} G_{tp} = d_p \qquad \forall p \in \mathcal{P} 
\end{equation}

Constraint~\ref{eq:max_simultaneous} limits the number of projects that may be active in any time period: 
\begin{equation}\label{eq:max_simultaneous} 
\sum_{p \in \mathcal{P}} G_{tp} \leq m \qquad \forall t \in \mathcal{T} 
\end{equation}

Constraint~\ref{eq:budget} ensures that the cumulative budget is not exceeded at any point in the planning horizon: 
\begin{equation}\label{eq:budget} 
\sum_{t = 0}^k \sum_{p \in \mathcal{P}} X_{tp} c_p \leq \sum_{t = 0}^k b_t \qquad \forall k \in \mathcal{T} 
\end{equation}

Finally, constraint~\ref{eq:domains} defines the binary nature of the decision variables: 
\begin{equation}\label{eq:domains} 
X_{tp}, G_{tp} \in \{0, 1\} \qquad \forall p \in \mathcal{P}, \forall t \in \mathcal{T} 
\end{equation}

\newpage
\section{Lower level Problem Formulation}
\label{appendix:LL_formulation}

This appendix provides the complete notation and constraints of the non-linear convex mathematical program for the lower level traffic assignment problem introduced in Section~\ref{subsec:LLproblem}.

\begin{table*}[h!]
\centering
\caption{Main notation for the lower level problem}
\label{table:LL_notation}
\begin{tabularx}{\textwidth}{p{3cm}X}
\toprule
\multicolumn{2}{l}{\textbf{Traffic Network: $\mathcal{G}=(\mathcal{N}, \mathcal{A})$}} \\
$n \in \mathcal{N}$ & Set of nodes \\
$\mathcal{W} \subseteq \mathcal{N} \times \mathcal{N}$ & Set of origin-destination (O-D) pairs \\
$m_w$ & Travel demand between O-D pair $w \in \mathcal{W}$ \\
$(i,j) \in \mathcal{A}$ & Set of links \\
\multicolumn{2}{l}{\textbf{Capacity and Travel Time Parameters}} \\
$\sigma_{ij}^{\text{base}}$ & Default road capacity on link $(i,j)$ \\
$\sigma_{ij}^{p}$ & Adjusted road capacity on link $(i,j)$ during project $p$ \\
$r_{ij}$ & Free-flow travel time on link $(i,j)$ \\
$r_{ij}^{p}$ & Adjusted free-flow travel time on link $(i,j)$ during project $p$ \\
\\
\multicolumn{2}{l}{\textbf{Path-Based Variables and Sets}} \\
$\Pi_w$ & Set of feasible paths between origin-destination pair $w \in \mathcal{W}$ \\
$x_\pi$ & Path flow on path $\pi \in \Pi_w$ \\
$c_\pi$ & Travel cost associated with path $\pi$ \\
$\lambda_w$ & Minimal path cost for OD pair $w$ (used in User Equillibrium conditions) \\
$\delta_{ij}^\pi$ & Incidence parameter: 1 if link $(i,j)$ is on path $\pi$, 0 otherwise \\
$f_{ij}$ & Total flow on link $(i,j)$ \\
\bottomrule
\end{tabularx}
\end{table*}

Traffic flow on all paths must be non-negative:
\begin{equation}
\label{eq:non-negative path flow}
    x_\pi \geq 0,\quad \forall \pi \in \Pi_w, \forall w \in \mathcal{W}
\end{equation}

The flow on a link equals the sum of flows on all paths that use that link:
\begin{equation}
\label{eq:link flow}
f_{ij} = \sum_{w \in \mathcal{W}} \sum_{\pi \in \Pi_w} \delta_{ij}^\pi \cdot x_\pi
\end{equation}

Path costs are computed as the sum of link costs along the path using incidence parameters:
\begin{equation}
\label{eq:path cost}
    c_\pi = \sum_{(i,j) \in \mathcal{A}} \delta_{ij}^\pi \cdot c_{ij}, \quad \forall \pi \in \Pi_w
\end{equation}

The total travel cost on each link follows the Bureau of Public Roads function~\citep{roadsTrafficAssignmentManual1964}, where $a$ and $b$ are empirical parameters commonly set at $a=0.15$ and $b=4$:
\begin{equation}
\label{eq:link cost}
    c_{ij} = r_{ij} \left(1 + a \left(\frac{f_{ij}}{\sigma_{ij}} \right)^b \right)
\end{equation}

When project $p$ is ongoing during period $t$, the capacity and free-flow travel time of the affected road segment $(i,j)$ are adjusted accordingly:
\begin{equation}
\label{eq:link capacity adjustment}
    \sigma_{ij}^{t}=
    \begin{cases}
        \sigma_{ij}^{\text{base}}, & \text{if } G_{tp}=0\\
        \sigma_{ij}^{p}, & \text{if } G_{tp}=1
    \end{cases} \quad \forall t, \forall (i,j) \in S
\end{equation}
\begin{equation}
\label{eq: link free-flow traveltime adjustment}
    r_{ij}^{t}=
    \begin{cases}
        r_{ij}^{\text{base}}, & \text{if } G_{tp}=0\\
        r_{ij}^{p}, & \text{if } G_{tp}=1
    \end{cases} \quad \forall t, \forall (i,j) \in S
\end{equation}

In our experiments, a road segment undergoing renovation has its capacity reduced to $\sigma_{ij}^{p} = 0.1\,\sigma_{ij}^{\text{base}}$ and its free-flow travel time increased to $r_{ij}^{p} = 1000\,r_{ij}^{\text{base}}$, which strongly discourages through-traffic on the affected segment while preserving network connectivity.

\newpage
\section{Frank-Wolfe Algorithm}
\label{appendix:frank_wolfe}

This appendix presents the pseudocode for the Conjugate Frank-Wolfe algorithm used in Section~\ref{subsec:llsm}, an accelerated variant of Frank-Wolfe that converges faster at equal accuracy by using conjugate search directions. This algorithm calculates the User-Equilibrium distribution of traffic flows across a road network. In our experiments the convergence tolerance is $\varepsilon = 0.01$, and the assignment is additionally capped at a maximum number of iterations.

\begin{algorithm}
    \caption{Conjugate Frank-Wolfe Algorithm for Traffic Assignment}
    \label{algo:TAP}
    \begin{algorithmic}[1]
        \State \textbf{Initialization:}
        \State For each OD pair $w \in W$, assign the demand $m_w$ to the shortest path based on free-flow travel times $r_{ij}$.
        \State Let $v^{ijt,0}_w$ denote the initial traffic flow on link $(i,j)$ toward destination $w$ at time $t$, and set iteration counter $k = 0$.
        
        \While{not converged}
            \State \textbf{Direction Finding (Shortest Path Calculation):}
            \State Update link travel costs $c_{ij}^{t,k}$ using the Bureau of Public Roads function:
            \[
                c_{ij}^{t,k} = r_{ij} \left(1 + \alpha \left(\frac{v^{ijt,k}_w}{\sigma_{ij}^t}\right)^\beta \right)
            \]
            \State For each $w \in W$, compute shortest paths from origin to destination based on $c_{ij}^{t,k}$.
            \State Determine the auxiliary flows $y^{ijt,k}_w$ via all-or-nothing assignment using the updated costs.

            \State \textbf{Conjugate Direction:}
            \State Combine the all-or-nothing direction with the previous search direction using a Polak--Ribi\`ere coefficient $\gamma^k$ to form a conjugate direction $d^{ijt,k}_w$, reverting to the all-or-nothing direction when conjugacy degrades (restart).

            \State \textbf{Line Search (Step Size Calculation):}
            \State Compute the optimal step size:
            \[
                \lambda^k = \arg\min_{\lambda \in [0,1]} f(v^{ijt,k}_w + \lambda (d^{ijt,k}_w - v^{ijt,k}_w))
            \]
            which minimizes the total travel cost objective across all links and OD pairs.

            \State \textbf{Flow Update:}
            \State Update traffic flows:
            \[
                v^{ijt,k+1}_w = v^{ijt,k}_w + \lambda^k (d^{ijt,k}_w - v^{ijt,k}_w)
            \]

            \State \textbf{Convergence Check:}
            \State Compute the relative gap $\text{gap}^k = \text{TSTT}^k / \text{SPTT}^k - 1$, where $\text{TSTT}^k$ is the total system travel time at the current flows and $\text{SPTT}^k$ is the travel time of the all-or-nothing assignment at the current costs.
            \If{$\text{gap}^k \leq \varepsilon$}
                \State Terminate the algorithm.
            \Else
                \State Increment $k \leftarrow k+1$ and return to Step 4.
            \EndIf
        \EndWhile
    \end{algorithmic}
\end{algorithm}

\newpage
\section{CostliestSubsetHeuristic}
\label{appendix:heuristic}

This appendix describes the CostliestSubsetHeuristic surrogate model used in Section~\ref{subsubsec:PLBE}.

The CostliestSubsetHeuristic is designed to exploit the empirical observation that traffic disruption generally increases as more roads are removed from the network. Given an input $\{p\}_{i,t}$---the set of ongoing renovation projects for solution $i$ at time $t$---the heuristic searches the set of previously evaluated simulations $\mathcal{S}^{\text{known}}$ for all subsets of $\{p\}_{i,t}$. It then returns the maximum $TTD$ value observed among these subsets, defaulting to $\{p\}=\emptyset$ (the empty network with no closures) if no subsets have been previously evaluated.

This approach provides a conservative lower bound estimate: since removing additional roads from an already-congested network typically increases travel delay, the congestion caused by closing roads $\{p\}_{i,t}$ should be at least as severe as the congestion caused by closing any subset of those roads. The heuristic requires no training and improves automatically as more simulations are performed during optimization, making it computationally inexpensive while providing reasonable estimates in the early stages of the algorithm when limited simulation data is available.

\newpage
\section{Network Variants: Topologies and Per-Seed Distributions}
\label{appendix:network_variants}

\begin{figure*}[htbp]
    \centering
    \includegraphics[width=\textwidth]{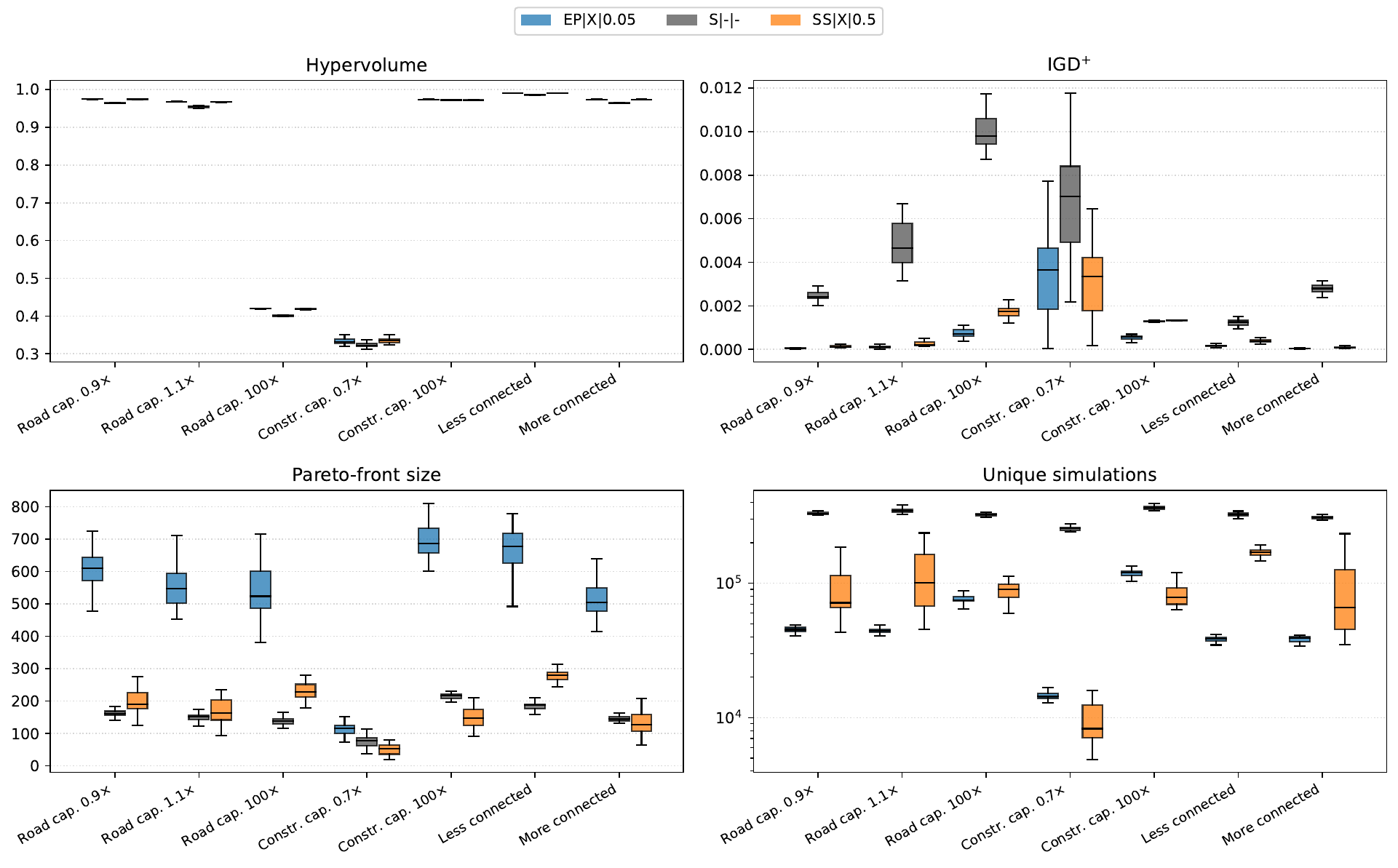}
    \caption{Experiment 3: per-seed distributions (30 seeds) of hypervolume, IGD$^{+}$, Pareto-front size, and unique simulations (log scale) for \texttt{EP|X|0.05}, standard NSGA-II (\texttt{S|-|-}), and the schedule-level SAEA baseline (\texttt{SS|X|0.5}) across the seven SF-76 variants. PLBE (blue) attains the lowest IGD$^{+}$ and the largest Pareto fronts using the fewest simulations in every variant; hypervolume saturates because each variant is scored against its own reference point.}
    \label{fig:e3_boxplots}
\end{figure*}

\begin{figure*}[htbp]
    \centering
    \begin{subfigure}[b]{0.48\textwidth}
        \centering
        \includegraphics[width=\textwidth]{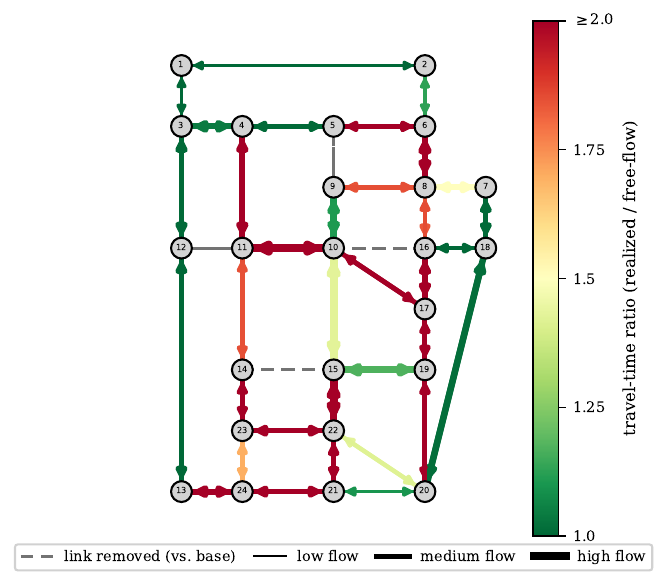}
        \caption{Less connected Sioux Falls network.}
        \label{fig:sioux_falls_less_connected}
    \end{subfigure}
    \hfill
    \begin{subfigure}[b]{0.48\textwidth}
        \centering
        \includegraphics[width=\textwidth]{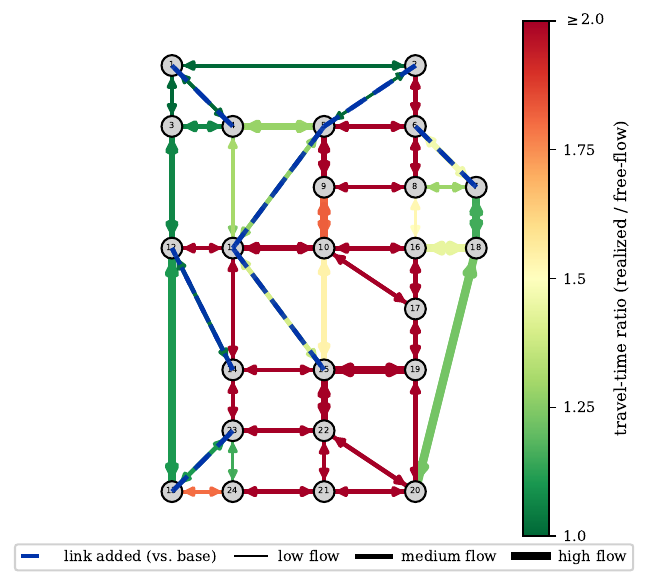}
        \caption{More connected Sioux Falls network.}
        \label{fig:sioux_falls_more_connected}
    \end{subfigure}
    \caption{Modified Sioux Falls network topologies used in Experiment 3.}
    \label{fig:network_variants}
\end{figure*}

Figure~\ref{fig:network_variants} shows the two topology-modified Sioux Falls networks used in the connectivity variants of Experiment~3 (Section~\ref{subsec:exp3_new}), and Figure~\ref{fig:e3_boxplots} the per-seed metric distributions underlying Table~\ref{tab:e3_results}.

\bibliographystyle{cas-model2-names}
\bibliography{references}

\begin{thebibliography}{53}
\expandafter\ifx\csname natexlab\endcsname\relax\def\natexlab#1{#1}\fi
\providecommand{\url}[1]{\texttt{#1}}
\providecommand{\href}[2]{#2}
\providecommand{\path}[1]{#1}
\providecommand{\DOIprefix}{doi:}
\providecommand{\ArXivprefix}{arXiv:}
\providecommand{\URLprefix}{URL: }
\providecommand{\Pubmedprefix}{pmid:}
\providecommand{\doi}[1]{\href{http://dx.doi.org/#1}{\path{#1}}}
\providecommand{\Pubmed}[1]{\href{pmid:#1}{\path{#1}}}
\providecommand{\bibinfo}[2]{#2}
\ifx\xfnm\relax \def\xfnm[#1]{\unskip,\space#1}\fi
%Type = Article
\bibitem[{Abdzadeh et~al.(2022)Abdzadeh, Noori and
  Ghannadpour}]{abdzadehSimultaneousSchedulingMultiple2022}
\bibinfo{author}{Abdzadeh, B.}, \bibinfo{author}{Noori, S.},
  \bibinfo{author}{Ghannadpour, S.F.}, \bibinfo{year}{2022}.
\newblock \bibinfo{title}{Simultaneous scheduling of multiple construction
  projects considering supplier selection and material transportation routing}.
\newblock \bibinfo{journal}{Automation in Construction} \bibinfo{volume}{140},
  \bibinfo{pages}{104336}.
\newblock \URLprefix
  \url{https://www.sciencedirect.com/science/article/pii/S0926580522002096},
  \DOIprefix\doi{10.1016/j.autcon.2022.104336}.
%Type = Article
\bibitem[{Aksoy et~al.(2021)Aksoy, Mutlu and Alver}]{aksoyUrbanRoadNetwork2021}
\bibinfo{author}{Aksoy, I.C.}, \bibinfo{author}{Mutlu, M.M.},
  \bibinfo{author}{Alver, Y.}, \bibinfo{year}{2021}.
\newblock \bibinfo{title}{Urban {Road} {Network} {Maintenance} {Scheduling}
  {Using} {Ant} {Colony} {Optimization}}.
\newblock \bibinfo{journal}{Gazi University Journal of Science}
  \bibinfo{volume}{34}, \bibinfo{pages}{710--716}.
\newblock \DOIprefix\doi{10.35378/gujs.789519}. \bibinfo{note}{number: 3}.
%Type = Article
\bibitem[{Alvanchi et~al.(2020)Alvanchi, Rahimi, Mousavi and
  Alikhani}]{alvanchiConstructionScheduleInfluential2020}
\bibinfo{author}{Alvanchi, A.}, \bibinfo{author}{Rahimi, M.},
  \bibinfo{author}{Mousavi, M.}, \bibinfo{author}{Alikhani, H.},
  \bibinfo{year}{2020}.
\newblock \bibinfo{title}{Construction schedule, an influential factor on air
  pollution in urban infrastructure projects}.
\newblock \bibinfo{journal}{Journal of Cleaner Production}
  \bibinfo{volume}{255}, \bibinfo{pages}{120222}.
\newblock \URLprefix
  \url{https://linkinghub.elsevier.com/retrieve/pii/S0959652620302699},
  \DOIprefix\doi{10.1016/j.jclepro.2020.120222}.
%Type = Article
\bibitem[{Asadifakhr et~al.(2025)Asadifakhr, Roy, Taherkhani, Han, Bell and
  Mo}]{asadifakhrMultiObjectiveGeneticAlgorithm2025}
\bibinfo{author}{Asadifakhr, K.}, \bibinfo{author}{Roy, S.G.},
  \bibinfo{author}{Taherkhani, A.H.}, \bibinfo{author}{Han, F.},
  \bibinfo{author}{Bell, E.S.}, \bibinfo{author}{Mo, W.}, \bibinfo{year}{2025}.
\newblock \bibinfo{title}{A {Multi}-{Objective} {Genetic} {Algorithm}
  {Approach} to {Sustainable} {Road}–{Stream} {Crossing} {Management}}.
\newblock \bibinfo{journal}{Sustainability} \bibinfo{volume}{17},
  \bibinfo{pages}{3987}.
\newblock \URLprefix \url{https://www.mdpi.com/2071-1050/17/9/3987},
  \DOIprefix\doi{10.3390/su17093987}.
%Type = Article
\bibitem[{Bagloee et~al.(2018a)Bagloee, Asadi, Sarvi and
  Patriksson}]{bagloeeHybridMachinelearningOptimization2018}
\bibinfo{author}{Bagloee, S.A.}, \bibinfo{author}{Asadi, M.},
  \bibinfo{author}{Sarvi, M.}, \bibinfo{author}{Patriksson, M.},
  \bibinfo{year}{2018}a.
\newblock \bibinfo{title}{A hybrid machine-learning and optimization method to
  solve bi-level problems}.
\newblock \bibinfo{journal}{Expert Systems with Applications}
  \bibinfo{volume}{95}, \bibinfo{pages}{142--152}.
\newblock \URLprefix
  \url{https://www.sciencedirect.com/science/article/pii/S0957417417307959},
  \DOIprefix\doi{10.1016/j.eswa.2017.11.039}.
%Type = Article
\bibitem[{Bagloee et~al.(2017)Bagloee, Sarvi and
  Patriksson}]{bagloeeHybridBranchandBoundBenders2017}
\bibinfo{author}{Bagloee, S.A.}, \bibinfo{author}{Sarvi, M.},
  \bibinfo{author}{Patriksson, M.}, \bibinfo{year}{2017}.
\newblock \bibinfo{title}{A {Hybrid} {Branch}‐and‐{Bound} and {Benders}
  {Decomposition} {Algorithm} for the {Network} {Design} {Problem}}.
\newblock \bibinfo{journal}{Computer-Aided Civil and Infrastructure
  Engineering} \bibinfo{volume}{32}, \bibinfo{pages}{319--343}.
\newblock \URLprefix
  \url{https://linkinghub.elsevier.com/retrieve/pii/S1093968726008315},
  \DOIprefix\doi{10.1111/mice.12224}.
%Type = Article
\bibitem[{Bagloee et~al.(2018b)Bagloee, Sarvi, Patriksson and
  Asadi}]{bagloeeOptimizationRoadsConstruction2018}
\bibinfo{author}{Bagloee, S.A.}, \bibinfo{author}{Sarvi, M.},
  \bibinfo{author}{Patriksson, M.}, \bibinfo{author}{Asadi, M.},
  \bibinfo{year}{2018}b.
\newblock \bibinfo{title}{Optimization for {Roads}' {Construction}:
  {Selection}, {Prioritization}, and {Scheduling}}.
\newblock \bibinfo{journal}{Computer-Aided Civil and Infrastructure
  Engineering} \bibinfo{volume}{33}, \bibinfo{pages}{833--848}.
\newblock \DOIprefix\doi{10.1111/mice.12370}.
%Type = Article
\bibitem[{Bai et~al.(2015)Bai, Ahmed, Li and
  Labi}]{baiHybridParetoFrontier2015}
\bibinfo{author}{Bai, Q.}, \bibinfo{author}{Ahmed, A.}, \bibinfo{author}{Li,
  Z.}, \bibinfo{author}{Labi, S.}, \bibinfo{year}{2015}.
\newblock \bibinfo{title}{A {Hybrid} {Pareto} {Frontier} {Generation} {Method}
  for {Trade}‐{Off} {Analysis} in {Transportation} {Asset} {Management}}.
\newblock \bibinfo{journal}{Computer-Aided Civil and Infrastructure
  Engineering} \bibinfo{volume}{30}, \bibinfo{pages}{163--180}.
\newblock \URLprefix
  \url{https://linkinghub.elsevier.com/retrieve/pii/S1093968726011783},
  \DOIprefix\doi{10.1111/mice.12079}.
%Type = Article
\bibitem[{Bai et~al.(2012)Bai, Labi and
  Sinha}]{baiTradeOffAnalysisMultiobjective2012}
\bibinfo{author}{Bai, Q.}, \bibinfo{author}{Labi, S.}, \bibinfo{author}{Sinha,
  K.C.}, \bibinfo{year}{2012}.
\newblock \bibinfo{title}{Trade-{Off} {Analysis} for {Multiobjective}
  {Optimization} in {Transportation} {Asset} {Management} by {Generating}
  {Pareto} {Frontiers} {Using} {Extreme} {Points} {Nondominated} {Sorting}
  {Genetic} {Algorithm} {II}}.
\newblock \bibinfo{journal}{Journal of Transportation Engineering}
  \bibinfo{volume}{138}, \bibinfo{pages}{798--808}.
\newblock \URLprefix
  \url{https://ascelibrary.org/doi/10.1061/%28ASCE%29TE.1943-5436.0000369},
  \DOIprefix\doi{10.1061/(ASCE)TE.1943-5436.0000369}.
%Type = Book
\bibitem[{Beckmann et~al.(1956)Beckmann, McGuire and
  Winsten}]{beckmannStudiesEconomicsTransportation1956}
\bibinfo{author}{Beckmann, M.}, \bibinfo{author}{McGuire, C.B.},
  \bibinfo{author}{Winsten, C.B.}, \bibinfo{year}{1956}.
\newblock \bibinfo{title}{Studies in the {Economics} of {Transportation}}.
\newblock \bibinfo{publisher}{Yale University Press}.
\newblock \URLprefix \url{https://trid.trb.org/View/91120}.
  \bibinfo{note}{number: 226 pp}.
%Type = Article
\bibitem[{Biondini and
  Frangopol(2016)}]{biondiniLifeCyclePerformanceDeteriorating2016}
\bibinfo{author}{Biondini, F.}, \bibinfo{author}{Frangopol, D.M.},
  \bibinfo{year}{2016}.
\newblock \bibinfo{title}{Life-{Cycle} {Performance} of {Deteriorating}
  {Structural} {Systems} under {Uncertainty}: {Review}}.
\newblock \bibinfo{journal}{Journal of Structural Engineering}
  \bibinfo{volume}{142}, \bibinfo{pages}{F4016001}.
\newblock \URLprefix \url{https://doi.org/10.1061/(ASCE)ST.1943-541X.0001544},
  \DOIprefix\doi{10.1061/(ASCE)ST.1943-541X.0001544}.
%Type = Misc
\bibitem[{Bosch et~al.(2025)Bosch, Heeswijk, Rogetzer and
  Mes}]{boschMachineLearningPredictions2025}
\bibinfo{author}{Bosch, R.}, \bibinfo{author}{Heeswijk, W.v.},
  \bibinfo{author}{Rogetzer, P.}, \bibinfo{author}{Mes, M.},
  \bibinfo{year}{2025}.
\newblock \bibinfo{title}{Machine {Learning} {Predictions} for {Traffic}
  {Equilibria} in {Road} {Renovation} {Scheduling}}.
\newblock \URLprefix \url{http://arxiv.org/abs/2506.05933},
  \DOIprefix\doi{10.48550/arXiv.2506.05933}. \bibinfo{note}{arXiv:2506.05933
  [cs]}.
%Type = Article
\bibitem[{Chai and Liang(2020)}]{chaiImprovedNSGAIIAlgorithm2020}
\bibinfo{author}{Chai, S.}, \bibinfo{author}{Liang, Q.}, \bibinfo{year}{2020}.
\newblock \bibinfo{title}{An {Improved} {NSGA}-{II} {Algorithm} for {Transit}
  {Network} {Design} and {Frequency} {Setting} {Problem}}.
\newblock \bibinfo{journal}{Journal of Advanced Transportation}
  \bibinfo{volume}{2020}, \bibinfo{pages}{1--20}.
\newblock \URLprefix \url{https://www.hindawi.com/journals/jat/2020/2895320/},
  \DOIprefix\doi{10.1155/2020/2895320}.
%Type = Inproceedings
\bibitem[{Chang et~al.(2001)Chang, Sawaya and
  Ziliaskopoulos}]{changTabuSearchBased}
\bibinfo{author}{Chang, Y.Y.}, \bibinfo{author}{Sawaya, O.B.},
  \bibinfo{author}{Ziliaskopoulos, A.}, \bibinfo{year}{2001}.
\newblock \bibinfo{title}{A {Tabu} {Search} {Based} {Approach} for {Work}
  {Zone} {Scheduling}}, in: \bibinfo{booktitle}{80th {Annual} {Meeting} of the
  {Transportation} {Research} {Board}}, \bibinfo{publisher}{Transportation
  Research Board}, \bibinfo{address}{Washington, D.C.}
\newblock \bibinfo{note}{Paper No. 01-2950}.
%Type = Inproceedings
\bibitem[{Chen and Guestrin(2016)}]{chenXGBoostScalableTree2016}
\bibinfo{author}{Chen, T.}, \bibinfo{author}{Guestrin, C.},
  \bibinfo{year}{2016}.
\newblock \bibinfo{title}{{XGBoost}: {A} {Scalable} {Tree} {Boosting}
  {System}}, in: \bibinfo{booktitle}{Proceedings of the 22nd {ACM} {SIGKDD}
  {International} {Conference} on {Knowledge} {Discovery} and {Data} {Mining}},
  \bibinfo{publisher}{Association for Computing Machinery},
  \bibinfo{address}{New York, NY, USA}. pp. \bibinfo{pages}{785--794}.
\newblock \URLprefix \url{https://dl.acm.org/doi/10.1145/2939672.2939785},
  \DOIprefix\doi{10.1145/2939672.2939785}.
%Type = Article
\bibitem[{Chen et~al.(2024)Chen, Zheng and
  Tan}]{chenRoadsideLiDARPlacement2024}
\bibinfo{author}{Chen, Y.}, \bibinfo{author}{Zheng, L.}, \bibinfo{author}{Tan,
  Z.}, \bibinfo{year}{2024}.
\newblock \bibinfo{title}{Roadside {LiDAR} placement for cooperative traffic
  detection by a novel chance constrained stochastic simulation optimization
  approach}.
\newblock \bibinfo{journal}{Transportation Research Part C: Emerging
  Technologies} \bibinfo{volume}{167}, \bibinfo{pages}{104838}.
\newblock \DOIprefix\doi{10.1016/j.trc.2024.104838}.
%Type = Article
\bibitem[{Cheu et~al.(2004)Cheu, Wang and
  Fwa}]{cheuGeneticAlgorithmSimulationMethodology2004}
\bibinfo{author}{Cheu, R.L.}, \bibinfo{author}{Wang, Y.}, \bibinfo{author}{Fwa,
  T.F.}, \bibinfo{year}{2004}.
\newblock \bibinfo{title}{Genetic {Algorithm}-{Simulation} {Methodology} for
  {Pavement} {Maintenance} {Scheduling}}.
\newblock \bibinfo{journal}{Computer-Aided Civil and Infrastructure
  Engineering} \bibinfo{volume}{19}, \bibinfo{pages}{446--455}.
\newblock \DOIprefix\doi{10.1111/j.1467-8667.2004.00369.x}.
%Type = Article
\bibitem[{Chien and Tang(2014)}]{chienSchedulingHighwayWork2014}
\bibinfo{author}{Chien, S.I.J.}, \bibinfo{author}{Tang, Y.},
  \bibinfo{year}{2014}.
\newblock \bibinfo{title}{Scheduling highway work zones with genetic algorithm
  considering the impact of traffic diversion}.
\newblock \bibinfo{journal}{Journal of Advanced Transportation}
  \bibinfo{volume}{48}, \bibinfo{pages}{287--303}.
\newblock \URLprefix
  \url{https://onlinelibrary.wiley.com/doi/abs/10.1002/atr.213},
  \DOIprefix\doi{10.1002/atr.213}.
%Type = Article
\bibitem[{Chowdhury et~al.(2025)Chowdhury, Guler and
  Stoffels}]{chowdhuryMultiobjectiveOptimizationMaintenance2025}
\bibinfo{author}{Chowdhury, L.F.}, \bibinfo{author}{Guler, S.I.},
  \bibinfo{author}{Stoffels, S.M.}, \bibinfo{year}{2025}.
\newblock \bibinfo{title}{Multi-objective optimization of maintenance, repair,
  and rehabilitation schedule for pavements in urban networks considering
  greenhouse gas emissions}.
\newblock \bibinfo{journal}{Structure and Infrastructure Engineering} ,
  \bibinfo{pages}{1--15}\URLprefix
  \url{https://www.tandfonline.com/doi/full/10.1080/15732479.2025.2565648},
  \DOIprefix\doi{10.1080/15732479.2025.2565648}.
%Type = Article
\bibitem[{Deb et~al.(2002)Deb, Pratap, Agarwal and
  Meyarivan}]{debFastElitistMultiobjective2002}
\bibinfo{author}{Deb, K.}, \bibinfo{author}{Pratap, A.},
  \bibinfo{author}{Agarwal, S.}, \bibinfo{author}{Meyarivan, T.},
  \bibinfo{year}{2002}.
\newblock \bibinfo{title}{A fast and elitist multiobjective genetic algorithm:
  {NSGA}-{II}}.
\newblock \bibinfo{journal}{IEEE Transactions on Evolutionary Computation}
  \bibinfo{volume}{6}, \bibinfo{pages}{182--197}.
\newblock \URLprefix
  \url{https://ieeexplore.ieee.org/document/996017/?arnumber=996017},
  \DOIprefix\doi{10.1109/4235.996017}. \bibinfo{note}{conference Name: IEEE
  Transactions on Evolutionary Computation}.
%Type = Article
\bibitem[{Derrac et~al.(2011)Derrac, García, Molina and
  Herrera}]{derracPracticalTutorialUse2011}
\bibinfo{author}{Derrac, J.}, \bibinfo{author}{García, S.},
  \bibinfo{author}{Molina, D.}, \bibinfo{author}{Herrera, F.},
  \bibinfo{year}{2011}.
\newblock \bibinfo{title}{A practical tutorial on the use of nonparametric
  statistical tests as a methodology for comparing evolutionary and swarm
  intelligence algorithms}.
\newblock \bibinfo{journal}{Swarm and Evolutionary Computation}
  \bibinfo{volume}{1}, \bibinfo{pages}{3--18}.
\newblock \URLprefix
  \url{https://linkinghub.elsevier.com/retrieve/pii/S2210650211000034},
  \DOIprefix\doi{10.1016/j.swevo.2011.02.002}.
%Type = Article
\bibitem[{Farahani et~al.(2013)Farahani, Miandoabchi, Szeto and
  Rashidi}]{farahaniReviewUrbanTransportation2013}
\bibinfo{author}{Farahani, R.Z.}, \bibinfo{author}{Miandoabchi, E.},
  \bibinfo{author}{Szeto, W.Y.}, \bibinfo{author}{Rashidi, H.},
  \bibinfo{year}{2013}.
\newblock \bibinfo{title}{A review of urban transportation network design
  problems}.
\newblock \bibinfo{journal}{European Journal of Operational Research}
  \bibinfo{volume}{229}, \bibinfo{pages}{281--302}.
\newblock \URLprefix
  \url{https://www.sciencedirect.com/science/article/pii/S0377221713000106},
  \DOIprefix\doi{10.1016/j.ejor.2013.01.001}.
%Type = Article
\bibitem[{Fuggini et~al.(2013)Fuggini, Chatzi and
  Zangani}]{fugginiCombiningGeneticAlgorithms2013}
\bibinfo{author}{Fuggini, C.}, \bibinfo{author}{Chatzi, E.},
  \bibinfo{author}{Zangani, D.}, \bibinfo{year}{2013}.
\newblock \bibinfo{title}{Combining {Genetic} {Algorithms} with a
  {Meso}‐{Scale} {Approach} for {System} {Identification} of a {Smart}
  {Polymeric} {Textile}}.
\newblock \bibinfo{journal}{Computer-Aided Civil and Infrastructure
  Engineering} \bibinfo{volume}{28}, \bibinfo{pages}{227--245}.
\newblock \URLprefix
  \url{https://linkinghub.elsevier.com/retrieve/pii/S1093968726008935},
  \DOIprefix\doi{10.1111/j.1467-8667.2012.00789.x}.
%Type = Article
\bibitem[{Gong and Fan(2016)}]{gongOptimizingSchedulingLongterm2016}
\bibinfo{author}{Gong, L.}, \bibinfo{author}{Fan, W.}, \bibinfo{year}{2016}.
\newblock \bibinfo{title}{Optimizing scheduling of long-term highway work zone
  projects}.
\newblock \bibinfo{journal}{International Journal of Transportation Science and
  Technology} \bibinfo{volume}{5}, \bibinfo{pages}{17--27}.
\newblock \DOIprefix\doi{10.1016/j.ijtst.2016.06.003}.
%Type = Article
\bibitem[{Hosseininasab and
  Shetab-Boushehri(2015)}]{hosseininasabIntegrationSelectingScheduling2015}
\bibinfo{author}{Hosseininasab, S.M.}, \bibinfo{author}{Shetab-Boushehri,
  S.N.}, \bibinfo{year}{2015}.
\newblock \bibinfo{title}{Integration of selecting and scheduling urban road
  construction projects as a time-dependent discrete network design problem}.
\newblock \bibinfo{journal}{European Journal of Operational Research}
  \bibinfo{volume}{246}, \bibinfo{pages}{762--771}.
\newblock \DOIprefix\doi{10.1016/j.ejor.2015.05.039}.
%Type = Article
\bibitem[{Hosseininasab et~al.(2018)Hosseininasab, Shetab-Boushehri, Hejazi and
  Karimi}]{hosseininasabMultiobjectiveIntegratedModel2018}
\bibinfo{author}{Hosseininasab, S.M.}, \bibinfo{author}{Shetab-Boushehri,
  S.N.}, \bibinfo{author}{Hejazi, S.}, \bibinfo{author}{Karimi, H.},
  \bibinfo{year}{2018}.
\newblock \bibinfo{title}{A multi-objective integrated model for selecting,
  scheduling, and budgeting road construction projects}.
\newblock \bibinfo{journal}{European Journal of Operational Research}
  \bibinfo{volume}{271}, \bibinfo{pages}{262--277}.
\newblock \DOIprefix\doi{10.1016/j.ejor.2018.04.051}.
%Type = Article
\bibitem[{Jiang and Szeto(2015)}]{jiangTimedependentTransportationNetwork2015}
\bibinfo{author}{Jiang, Y.}, \bibinfo{author}{Szeto, W.}, \bibinfo{year}{2015}.
\newblock \bibinfo{title}{Time-dependent transportation network design that
  considers health cost}.
\newblock \bibinfo{journal}{Transportmetrica A: Transport Science}
  \bibinfo{volume}{11}, \bibinfo{pages}{74--101}.
\newblock \URLprefix \url{https://doi.org/10.1080/23249935.2014.927938},
  \DOIprefix\doi{10.1080/23249935.2014.927938}.
%Type = Article
\bibitem[{Katoch et~al.(2021)Katoch, Chauhan and
  Kumar}]{katochReviewGeneticAlgorithm2021}
\bibinfo{author}{Katoch, S.}, \bibinfo{author}{Chauhan, S.S.},
  \bibinfo{author}{Kumar, V.}, \bibinfo{year}{2021}.
\newblock \bibinfo{title}{A review on genetic algorithm: past, present, and
  future}.
\newblock \bibinfo{journal}{Multimedia Tools and Applications}
  \bibinfo{volume}{80}, \bibinfo{pages}{8091--8126}.
\newblock \URLprefix \url{https://doi.org/10.1007/s11042-020-10139-6},
  \DOIprefix\doi{10.1007/s11042-020-10139-6}.
%Type = Article
\bibitem[{Kim et~al.(2024)Kim, Butt, Harvey and
  Ostovar}]{kimEnvironmentalImpactsTraffic2024}
\bibinfo{author}{Kim, C.}, \bibinfo{author}{Butt, A.A.},
  \bibinfo{author}{Harvey, J.T.}, \bibinfo{author}{Ostovar, M.},
  \bibinfo{year}{2024}.
\newblock \bibinfo{title}{Environmental impacts from traffic on highway
  construction work zones: {Framework} and simulations}.
\newblock \bibinfo{journal}{International Journal of Sustainable
  Transportation} \bibinfo{volume}{18}, \bibinfo{pages}{680--694}.
\newblock \URLprefix
  \url{https://www.tandfonline.com/doi/full/10.1080/15568318.2024.2392624},
  \DOIprefix\doi{10.1080/15568318.2024.2392624}.
%Type = Article
\bibitem[{Kumar and Mishra(2018)}]{kumarSimplifiedFrameworkSequencing2018}
\bibinfo{author}{Kumar, A.}, \bibinfo{author}{Mishra, S.},
  \bibinfo{year}{2018}.
\newblock \bibinfo{title}{A simplified framework for sequencing of
  transportation projects considering user costs and benefits}.
\newblock \bibinfo{journal}{Transportmetrica A: Transport Science}
  \bibinfo{volume}{14}, \bibinfo{pages}{346--371}.
\newblock \DOIprefix\doi{10.1080/23249935.2017.1387827}.
%Type = Article
\bibitem[{Lee(2009)}]{leeOptimizingScheduleImproving2009}
\bibinfo{author}{Lee, H.Y.}, \bibinfo{year}{2009}.
\newblock \bibinfo{title}{Optimizing schedule for improving the traffic impact
  of work zone on roads}.
\newblock \bibinfo{journal}{Automation in Construction} \bibinfo{volume}{18},
  \bibinfo{pages}{1034--1044}.
\newblock \URLprefix
  \url{https://www.sciencedirect.com/science/article/pii/S0926580509000867},
  \DOIprefix\doi{10.1016/j.autcon.2009.05.004}.
%Type = Article
\bibitem[{Li and Fan(2021)}]{liBilevelOptimizationLongterm2021}
\bibinfo{author}{Li, Y.}, \bibinfo{author}{Fan, W.}, \bibinfo{year}{2021}.
\newblock \bibinfo{title}{Bi-level optimization of long-term highway work zone
  scheduling considering elastic demand}.
\newblock \bibinfo{journal}{Smart and Resilient Transportation}
  \bibinfo{volume}{3}, \bibinfo{pages}{118--130}.
\newblock \DOIprefix\doi{10.1108/SRT-01-2021-0004}.
%Type = Article
\bibitem[{Lu et~al.(2016)Lu, Xiong, Ding and Wang}]{luOptimalScheduleUrban2016}
\bibinfo{author}{Lu, G.}, \bibinfo{author}{Xiong, Y.}, \bibinfo{author}{Ding,
  C.}, \bibinfo{author}{Wang, Y.}, \bibinfo{year}{2016}.
\newblock \bibinfo{title}{An {Optimal} {Schedule} for {Urban} {Road} {Network}
  {Repair} {Based} on the {Greedy} {Algorithm}}.
\newblock \bibinfo{journal}{PLOS ONE} \bibinfo{volume}{11},
  \bibinfo{pages}{e0164780}.
\newblock \DOIprefix\doi{10.1371/journal.pone.0164780}.
%Type = Article
\bibitem[{Ma et~al.(2018)Ma, Cheng and Li}]{maRoadMaintenanceOptimization2018}
\bibinfo{author}{Ma, J.}, \bibinfo{author}{Cheng, L.}, \bibinfo{author}{Li,
  D.}, \bibinfo{year}{2018}.
\newblock \bibinfo{title}{Road {Maintenance} {Optimization} {Model} {Based} on
  {Dynamic} {Programming} in {Urban} {Traffic} {Network}}.
\newblock \bibinfo{journal}{Journal of Advanced Transportation}
  \bibinfo{volume}{2018}, \bibinfo{pages}{4539324}.
\newblock \URLprefix
  \url{https://onlinelibrary.wiley.com/doi/abs/10.1155/2018/4539324},
  \DOIprefix\doi{10.1155/2018/4539324}.
%Type = Article
\bibitem[{Miralinaghi et~al.(2022)Miralinaghi, Davatgari, Seilabi and
  Labi}]{miralinaghiContractBundlingConsiderations2022}
\bibinfo{author}{Miralinaghi, M.}, \bibinfo{author}{Davatgari, A.},
  \bibinfo{author}{Seilabi, S.E.}, \bibinfo{author}{Labi, S.},
  \bibinfo{year}{2022}.
\newblock \bibinfo{title}{Contract bundling considerations in urban road
  project scheduling}.
\newblock \bibinfo{journal}{Computer-Aided Civil and Infrastructure
  Engineering} \bibinfo{volume}{37}, \bibinfo{pages}{427--450}.
\newblock \DOIprefix\doi{10.1111/mice.12740}.
%Type = Article
\bibitem[{Miralinaghi et~al.(2020a)Miralinaghi, Seilabi, Chen, Hsu and
  Labi}]{miralinaghiOptimizingSelectionScheduling2020}
\bibinfo{author}{Miralinaghi, M.}, \bibinfo{author}{Seilabi, S.},
  \bibinfo{author}{Chen, S.}, \bibinfo{author}{Hsu, Y.T.},
  \bibinfo{author}{Labi, S.}, \bibinfo{year}{2020}a.
\newblock \bibinfo{title}{Optimizing the selection and scheduling of
  multi-class projects using a {Stackelberg} framework}.
\newblock \bibinfo{journal}{European Journal of Operational Research}
  \bibinfo{volume}{286}, \bibinfo{pages}{508--522}.
\newblock \DOIprefix\doi{10.1016/j.ejor.2020.03.051}.
%Type = Article
\bibitem[{Miralinaghi et~al.(2020b)Miralinaghi, Woldemariam, Abraham, Chen,
  Labi and Chen}]{miralinaghiNetworklevelSchedulingRoad2020}
\bibinfo{author}{Miralinaghi, M.}, \bibinfo{author}{Woldemariam, W.},
  \bibinfo{author}{Abraham, D.M.}, \bibinfo{author}{Chen, S.},
  \bibinfo{author}{Labi, S.}, \bibinfo{author}{Chen, Z.},
  \bibinfo{year}{2020}b.
\newblock \bibinfo{title}{Network-level scheduling of road construction
  projects considering user and business impacts}.
\newblock \bibinfo{journal}{Computer-Aided Civil and Infrastructure
  Engineering} \bibinfo{volume}{35}, \bibinfo{pages}{650--667}.
\newblock \DOIprefix\doi{10.1111/mice.12518}.
%Type = Article
\bibitem[{Moghtadernejad et~al.(2022)Moghtadernejad, Adey and
  Hackl}]{moghtadernejadPrioritizingRoadNetwork2022}
\bibinfo{author}{Moghtadernejad, S.}, \bibinfo{author}{Adey, B.T.},
  \bibinfo{author}{Hackl, J.}, \bibinfo{year}{2022}.
\newblock \bibinfo{title}{Prioritizing {Road} {Network} {Restorative}
  {Interventions} {Using} a {Discrete} {Particle} {Swarm} {Optimization}}.
\newblock \bibinfo{journal}{Journal of Infrastructure Systems}
  \bibinfo{volume}{28}, \bibinfo{pages}{04022039}.
\newblock \URLprefix
  \url{https://ascelibrary.org/doi/10.1061/(ASCE)IS.1943-555X.0000725},
  \DOIprefix\doi{10.1061/(ASCE)IS.1943-555X.0000725}.
%Type = Article
\bibitem[{Nguyen(1974)}]{nguyenAlgorithmTrafficAssignment1974}
\bibinfo{author}{Nguyen, S.}, \bibinfo{year}{1974}.
\newblock \bibinfo{title}{An {Algorithm} for the {Traffic} {Assignment}
  {Problem}}.
\newblock \bibinfo{journal}{Transportation Science} \bibinfo{volume}{8},
  \bibinfo{pages}{203--216}.
\newblock \URLprefix \url{https://www.jstor.org/stable/25767747}.
%Type = Misc
\bibitem[{NOS(2020)}]{nosKademuurCentrumAmsterdam2020}
\bibinfo{author}{NOS}, \bibinfo{year}{2020}.
\newblock \bibinfo{title}{Kademuur in centrum {Amsterdam} deels ingestort}.
\newblock \URLprefix
  \url{https://nos.nl/artikel/2346286-kademuur-in-centrum-amsterdam-deels-ingestort}.
%Type = Article
\bibitem[{Poorzahedy and
  Abulghasemi(2005)}]{poorzahedyApplicationAntSystem2005}
\bibinfo{author}{Poorzahedy, H.}, \bibinfo{author}{Abulghasemi, F.},
  \bibinfo{year}{2005}.
\newblock \bibinfo{title}{Application of {Ant} {System} to network design
  problem}.
\newblock \bibinfo{journal}{Transportation} \bibinfo{volume}{32},
  \bibinfo{pages}{251--273}.
\newblock \URLprefix \url{https://doi.org/10.1007/s11116-004-8246-7},
  \DOIprefix\doi{10.1007/s11116-004-8246-7}.
%Type = Book
\bibitem[{Roads(1964)}]{roadsTrafficAssignmentManual1964}
\bibinfo{author}{Roads, U.S.B.o.P.}, \bibinfo{year}{1964}.
\newblock \bibinfo{title}{Traffic {Assignment} {Manual} for {Application} with
  a {Large}, {High} {Speed} {Computer}}.
\newblock \bibinfo{publisher}{U.S. Department of Commerce, Bureau of Public
  Roads, Office of Planning, Urban Planning Division}.
\newblock \bibinfo{note}{Google-Books-ID: AvNUR\_O\_JEcC}.
%Type = Article
\bibitem[{Ruiz-Vélez et~al.(2024)Ruiz-Vélez, García, Alcalá and
  Yepes}]{ruiz-velezSustainableRoadInfrastructure2024}
\bibinfo{author}{Ruiz-Vélez, A.}, \bibinfo{author}{García, J.},
  \bibinfo{author}{Alcalá, J.}, \bibinfo{author}{Yepes, V.},
  \bibinfo{year}{2024}.
\newblock \bibinfo{title}{Sustainable {Road} {Infrastructure}
  {Decision}-{Making}: {Custom} {NSGA}-{II} with {Repair} {Operators} for
  {Multi}-{Objective} {Optimization}}.
\newblock \bibinfo{journal}{Mathematics} \bibinfo{volume}{12},
  \bibinfo{pages}{730}.
\newblock \URLprefix \url{https://www.mdpi.com/2227-7390/12/5/730},
  \DOIprefix\doi{10.3390/math12050730}.
%Type = Article
\bibitem[{Saneii et~al.(2023)Saneii, Kazemeini, Seilabi, Miralinaghi and
  Labi}]{saneiiMethodologySchedulingWithinday2023}
\bibinfo{author}{Saneii, M.}, \bibinfo{author}{Kazemeini, A.},
  \bibinfo{author}{Seilabi, S.E.}, \bibinfo{author}{Miralinaghi, M.},
  \bibinfo{author}{Labi, S.}, \bibinfo{year}{2023}.
\newblock \bibinfo{title}{A methodology for scheduling within‐day roadway
  work zones using deep neural networks and active learning}.
\newblock \bibinfo{journal}{Computer-Aided Civil and Infrastructure
  Engineering} \bibinfo{volume}{38}, \bibinfo{pages}{1101--1126}.
\newblock \URLprefix
  \url{https://linkinghub.elsevier.com/retrieve/pii/S1093968726014647},
  \DOIprefix\doi{10.1111/mice.12921}.
%Type = Article
\bibitem[{Seilabi et~al.(2025)Seilabi, Saneii, Pourgholamali, Miralinaghi and
  Labi}]{seilabiReinforcementLearningbasedApproach2025}
\bibinfo{author}{Seilabi, S.E.}, \bibinfo{author}{Saneii, M.},
  \bibinfo{author}{Pourgholamali, M.}, \bibinfo{author}{Miralinaghi, M.},
  \bibinfo{author}{Labi, S.}, \bibinfo{year}{2025}.
\newblock \bibinfo{title}{Reinforcement learning‐based approach for urban
  road project scheduling considering alternative closure types}.
\newblock \bibinfo{journal}{Comput.-Aided Civ. Infrastruct. Eng.}
  \bibinfo{volume}{40}, \bibinfo{pages}{721--740}.
\newblock \URLprefix \url{https://doi.org/10.1111/mice.13365},
  \DOIprefix\doi{10.1111/mice.13365}.
%Type = Techreport
\bibitem[{Song et~al.(2018)Song, He and
  Liu}]{songRehabilitationProjectSelection2018}
\bibinfo{author}{Song, Z.}, \bibinfo{author}{He, Y.}, \bibinfo{author}{Liu,
  Z.}, \bibinfo{year}{2018}.
\newblock \bibinfo{title}{Rehabilitation {Project} {Selection} and {Scheduling}
  in {Transportation} {Networks}}.
\newblock \bibinfo{type}{Technical Report} \bibinfo{number}{MPC-18-358}.
  Mountain Plains Consortium.
\newblock \URLprefix \url{https://rosap.ntl.bts.gov/view/dot/42564}.
%Type = Article
\bibitem[{Szeto et~al.(2010)Szeto, Jaber and
  O’Mahony}]{szetoTimeDependentDiscreteNetwork2010}
\bibinfo{author}{Szeto, W.Y.}, \bibinfo{author}{Jaber, X.},
  \bibinfo{author}{O’Mahony, M.}, \bibinfo{year}{2010}.
\newblock \bibinfo{title}{Time-{Dependent} {Discrete} {Network} {Design}
  {Frameworks} {Considering} {Land} {Use}}.
\newblock \bibinfo{journal}{Computer-Aided Civil and Infrastructure
  Engineering} \bibinfo{volume}{25}, \bibinfo{pages}{411--426}.
\newblock \DOIprefix\doi{10.1111/j.1467-8667.2010.00654.x}.
%Type = Article
\bibitem[{Tao and Schonfeld(2007)}]{taoIslandModelsStochastic2007}
\bibinfo{author}{Tao, X.}, \bibinfo{author}{Schonfeld, P.},
  \bibinfo{year}{2007}.
\newblock \bibinfo{title}{Island {Models} for {Stochastic} {Problem} of
  {Transportation} {Project} {Selection} and {Scheduling}}.
\newblock \bibinfo{journal}{Transportation Research Record}
  \bibinfo{volume}{2039}, \bibinfo{pages}{16--23}.
\newblock \URLprefix \url{https://doi.org/10.3141/2039-02},
  \DOIprefix\doi{10.3141/2039-02}.
%Type = Article
\bibitem[{Ukkusuri and
  Patil(2009)}]{ukkusuriMultiperiodTransportationNetwork2009}
\bibinfo{author}{Ukkusuri, S.V.}, \bibinfo{author}{Patil, G.},
  \bibinfo{year}{2009}.
\newblock \bibinfo{title}{Multi-period transportation network design under
  demand uncertainty}.
\newblock \bibinfo{journal}{Transportation Research Part B: Methodological}
  \bibinfo{volume}{43}, \bibinfo{pages}{625--642}.
\newblock \URLprefix
  \url{https://www.sciencedirect.com/science/article/pii/S0191261509000150},
  \DOIprefix\doi{10.1016/j.trb.2009.01.004}.
%Type = Article
\bibitem[{Xu et~al.(2009)Xu, Wei and Hu}]{xuStudyContinuousNetwork2009}
\bibinfo{author}{Xu, T.}, \bibinfo{author}{Wei, H.}, \bibinfo{author}{Hu, G.},
  \bibinfo{year}{2009}.
\newblock \bibinfo{title}{Study on continuous network design problem using
  simulated annealing and genetic algorithm}.
\newblock \bibinfo{journal}{Expert Systems with Applications}
  \bibinfo{volume}{36}, \bibinfo{pages}{1322--1328}.
\newblock \URLprefix
  \url{https://www.sciencedirect.com/science/article/pii/S0957417407005702},
  \DOIprefix\doi{10.1016/j.eswa.2007.11.023}.
%Type = Article
\bibitem[{Yamany et~al.(2024)Yamany, Cawley, Reza and
  Ksaibati}]{yamanyNetworklevelPavementMaintenance2024}
\bibinfo{author}{Yamany, M.S.}, \bibinfo{author}{Cawley, L.},
  \bibinfo{author}{Reza, I.}, \bibinfo{author}{Ksaibati, K.},
  \bibinfo{year}{2024}.
\newblock \bibinfo{title}{Network-level pavement maintenance and rehabilitation
  planning using genetic algorithm}.
\newblock \bibinfo{journal}{Innovative Infrastructure Solutions}
  \bibinfo{volume}{9}, \bibinfo{pages}{208}.
\newblock \URLprefix \url{https://doi.org/10.1007/s41062-024-01534-1},
  \DOIprefix\doi{10.1007/s41062-024-01534-1}.
%Type = Article
\bibitem[{Zhang et~al.(2024)Zhang, Li, Li, Zhang and
  Si}]{zhangOptimalConditionbasedMaintenance2024}
\bibinfo{author}{Zhang, Z.}, \bibinfo{author}{Li, H.}, \bibinfo{author}{Li,
  T.}, \bibinfo{author}{Zhang, J.}, \bibinfo{author}{Si, X.},
  \bibinfo{year}{2024}.
\newblock \bibinfo{title}{An optimal condition-based maintenance policy for
  nonlinear stochastic degrading systems}.
\newblock \bibinfo{journal}{Reliability Engineering \& System Safety}
  \bibinfo{volume}{251}, \bibinfo{pages}{110349}.
\newblock \URLprefix
  \url{https://www.sciencedirect.com/science/article/pii/S0951832024004216},
  \DOIprefix\doi{10.1016/j.ress.2024.110349}.
%Type = Article
\bibitem[{Zheng et~al.(2014)Zheng, Nava and
  Chiu}]{zhengMeasuringNetworkwideTraffic2014}
\bibinfo{author}{Zheng, H.}, \bibinfo{author}{Nava, E.}, \bibinfo{author}{Chiu,
  Y.C.}, \bibinfo{year}{2014}.
\newblock \bibinfo{title}{Measuring {Networkwide} {Traffic} {Delay} in
  {Schedule} {Optimization} for {Work}-{Zone} {Planning} in {Urban}
  {Networks}}.
\newblock \bibinfo{journal}{IEEE Transactions on Intelligent Transportation
  Systems} \bibinfo{volume}{15}, \bibinfo{pages}{2595--2604}.
\newblock \DOIprefix\doi{10.1109/TITS.2014.2318299}.

\end{thebibliography}

\end{document}